\documentclass[10pt,english,journal]{IEEEtran}
\usepackage[T1]{fontenc}
\usepackage[latin9]{inputenc}
\usepackage[active]{srcltx}
\usepackage{color}
\usepackage{babel}
\usepackage{array}
\usepackage{pifont}
\usepackage{float}
\usepackage{amsmath}
\usepackage{amsthm}
\usepackage{amssymb}
\usepackage{graphicx}
\usepackage[unicode=true,
 bookmarks=false,
 breaklinks=false,pdfborder={0 0 1},backref=false,colorlinks=false]
 {hyperref}
\hypersetup{pdftitle={Your Title},
 pdfauthor={Your Name},
 pdfborderstyle=,draft=True,pdfpagelayout=OneColumn,pdfnewwindow=true,pdfstartview=XYZ,plainpages=True}

\makeatletter

\newcommand{\noun}[1]{\textsc{#1}}
\newcommand{\lyxmathsym}[1]{\ifmmode\begingroup\def\b@ld{bold}
  \text{\ifx\math@version\b@ld\bfseries\fi#1}\endgroup\else#1\fi}

\providecommand{\tabularnewline}{\\}
\floatstyle{ruled}
\newfloat{algorithm}{tbp}{loa}
\providecommand{\algorithmname}{Algorithm}
\floatname{algorithm}{\protect\algorithmname}

\theoremstyle{plain}
\newtheorem{thm}{\protect\theoremname}
\theoremstyle{plain}
\newtheorem{prop}{\protect\propositionname}
\theoremstyle{remark}
\newtheorem{claim}{\protect\claimname}
\theoremstyle{definition}
\newtheorem{defn}{\protect\definitionname}

\usepackage{babel}
\usepackage{algpseudocode}

\usepackage[para]{footmisc}

\algdef{SE}{Begin}{End}{\textbf{begin}}{\textbf{end}}
\algnewcommand\algorithmicforeach{\textbf{for each}}
\algdef{S}[FOR]{ForEach}[1]{\algorithmicforeach\ #1\ \algorithmicdo}
\newcommand{\Break}{\State \textbf{break}}

\providecommand{\claimname}{Claim}
\providecommand{\definitionname}{Definition}
\providecommand{\propositionname}{Proposition}
\providecommand{\theoremname}{Theorem}

\@ifundefined{showcaptionsetup}{}{%
 \PassOptionsToPackage{caption=false}{subfig}}
\usepackage{subfig}
\makeatother

\providecommand{\claimname}{Claim}
\providecommand{\definitionname}{Definition}
\providecommand{\propositionname}{Proposition}
\providecommand{\theoremname}{Theorem}

\begin{document}
\title{Auction Mechanisms in Cloud/Fog Computing Resource Allocation for
Public Blockchain Networks }
\author{Yutao Jiao, Ping Wang, \IEEEmembership{Senior Member,~IEEE,} Dusit
Niyato, \IEEEmembership{Fellow,~IEEE,} and Kongrath Suankaewmanee\thanks{Yutao Jiao, Dusit Niyato and Kongrath Suankaewmanee are with the School
of Computer Science and Engineering, Nanyang Technological University,
Singapore. Ping Wang is with the Lassonde School of Engineering, York
University, Canada.}}

\maketitle

\begin{abstract}
As an emerging decentralized secure data management platform, blockchain
has gained much popularity recently. To maintain a canonical state
of blockchain data record, proof-of-work based consensus protocols
provide the nodes, referred to as miners, in the network with incentives
for confirming new block of transactions through a process of ``block
mining'' by solving a cryptographic puzzle. Under the circumstance
of limited local computing resources, e.g., mobile devices, it is
natural for rational miners, i.e., consensus nodes, to offload computational
tasks for proof of work to the cloud/fog computing servers. Therefore,
we focus on the trading between the cloud/fog computing service provider
and miners, and propose an auction-based market model for efficient
computing resource allocation. In particular, we consider a proof-of-work
based blockchain network, which is constrained by the computing resource
and deployed as an infrastructure for decentralized data management
applications. Due to the competition among miners in the blockchain
network, the allocative externalities are particularly taken into
account when designing the auction mechanisms. Specifically, we consider
two bidding schemes: the constant-demand scheme where each miner bids
for a fixed quantity of resources, and the multi-demand scheme where
the miners can submit their preferable demands and bids. For the constant-demand
bidding scheme, we propose an auction mechanism that achieves optimal
social welfare. In the multi-demand bidding scheme, the social welfare
maximization problem is NP-hard. Therefore, we design an approximate
algorithm which guarantees the truthfulness, individual rationality
and computational efficiency. Through extensive simulations, we show
that our proposed auction mechanisms with the two bidding schemes
can efficiently maximize the social welfare of the blockchain network
and provide effective strategies for the cloud/fog computing service
provider. 
\end{abstract}


\section{Introduction}

\IEEEPARstart{B}{y} contrast to traditional currencies, cryptocurrencies
are traded among participants over a peer-to-peer (P2P) network without
relying on third parties such as banks or financial regulatory authorities~\cite{Nakamoto2008}.
As the backbone technology of decentralized cryptocurrencies, blockchain
has also heralded many applications in various fields, such as finance~\cite{Guo2016},
Internet of Things (IoT)~\cite{Christidis2016} and resource offloading~\cite{Chatzopoulos2017}.
According to the market research firm Tractica's report, it is estimated
that the annual revenue for enterprise applications of blockchain
will increase to \$19.9 billion by 2025~\cite{Tractica}. Essentially,
blockchain is a tamperproof, distributed database that records transactional
data in a P2P network. The database state is decentrally maintained,
and any member node in the overlay blockchain network is permitted
to participate in the state maintenance without identity authentication.
The transactions among member nodes are recorded in cryptographic
hash-linked data structures known as \emph{blocks}. A series of confirmed
blocks are arranged in chronological order to form a sequential chain,
hence named \emph{blockchain}. All member nodes in the network are
required to follow the Nakamoto consensus protocol~\cite{Nakamoto2008}
(or other protocols alike), to agree on the transactional data, cryptographic
hashes and digital signatures stored in the block to guarantee the
integrity of the blockchain.

The Nakamoto consensus protocol integrates a critical computing-intensive
process, called \emph{Proof-of-Work} \emph{(PoW)}. In order to have
their local views of the blockchain accepted by the network as the
canonical state of the blockchain, consensus nodes (i.e., block miners)
have to solve a cryptographic puzzle, i.e., find a nonce to be contained
in the block such that the hash value of the entire block is smaller
than a preset target. This computational process is called \emph{mining},
where the consensus nodes which contribute their computing power to
mining are known as \emph{miners}. Typically, the mining task for
PoW can be regarded as a tournament~\cite{Garay2015}. First, each
miner collects and verifies a certain number of unconfirmed transaction
records which are aggregated into a new block. Next, all miners chase
each other to be the first one to obtain the desired nonce value as
the PoW solutions for the new block which combines the collected transactional
data\footnote{We refer to all transaction records stored in the block simply as
transactional data in the rest of the paper.} and block metadata. Once the PoW puzzle is solved, this new block
will be immediately broadcasted to the entire blockchain network.
Meanwhile, the other miners receive this message and perform a chain
validation-comparison process to decide whether to approve and add
newly generated block to the blockchain. The miner which successfully
has its proposed block linked to the blockchain will be given a certain
amount of reward, including a fixed bonus and a variable transaction
fee, as an incentive of mining.

Since no prior authorization is required, the permissionless blockchain
is especially suitable for serving as a platform for decentralized
autonomous data management in many applications. Some representative
examples can be found in data sharing~\cite{Shafagh2017}, electricity
trading in smart grid~\cite{Kang2017} and personal data access control~\cite{Zyskind2015}.
Apart from the feature of public access, permissionless blockchains
have the advantage in quickly establishing a self-organized data management
platform to support various decentralized applications (DApps). This
is a breakthrough in production relations in that people can independently
design smart contracts and freely build decentralized applications
themselves without the support or permission from trusted intermediaries.
By the PoW-based Nakamoto consensus protocol, people are encouraged
to become consensus nodes, i.e., miners, with the mining reward. \textcolor{black}{Unfortunately,
solving the PoW puzzle needs continuous, high computing power which
mobile devices and IoT devices cannot afford.} As the number of mobile
phone users is forecasted to reach nearly 5 billion\footnote{https://www.statista.com/statistics/274774/forecast-of-mobile- phone-users-worldwide}
in 2019, it is expected that DApps would usher in explosive growth
if mobile devices can join in the mining and consensus process and
self-organize a blockchain network to support DApps~\cite{Xiong2018a}.
To alleviate the computational bottleneck, the consensus nodes can
access the cloud/fog computing service to offload their mining tasks,
thus enabling blockchain-based DApps. As the cloud/fog computing service
can breed more consensus nodes to be able to execute the mining task,
it would significantly improve the robustness of the blockchain network
and then raise the valuation of DApps, which further attracts more
DApp users to join, forming a virtuous circle.

\textcolor{black}{In this paper, we mainly investigate the trading
between the cloud/fog computing service provider (CFP) and the computationally
lightweight devices, i.e., }\textcolor{black}{\emph{miners}}\textcolor{black}{.
From the system perspective, we aim to maximize the }\textcolor{black}{\emph{social
welfare}}\textcolor{black}{{} which is the total utility of the CFP
and all miners in the blockchain network.} The social welfare can
be interpreted as the system efficiency\textcolor{black}{~\cite{Zhang2017a}.}
For an efficient and sustainable business ecosystem, there are some
critical issues about cloud/fog resources allocation and pricing for
the service provider. First, which miner can be offered the computing
resources? Too many miners will cause service congestion and incur
high operation cost to the service provider. By contrast, a very small
group of miners may erode the integrity of the blockchain network.
Second, how to set a reasonable service price for miners such that
they can be incentivized to undertake the mining tasks? The efficient
method is to set up an auction where the miners can actively submit
their bids to the CFP for decision making. We should also consider
how to make miners truthfully expose their private valuation. A miner's
valuation on the computing service is directly related to its privately
collected transactional data which determines its expected reward
from the blockchain. To address the above questions, we propose an
auction-based cloud/fog computing resource market model for blockchain
networks. Moreover, we design truthful auction mechanisms for two
different bidding schemes. One is the \emph{constant-demand scheme}
where the CFP restricts that each miner can bid only for the same
quantity of computing resources. The other one is the \emph{multi-demand
scheme} where miners can request their demands and express the corresponding
bids more freely. The major contributions of this paper can be summarized
as follows: 
\begin{itemize}
\item In the auction-based cloud/fog computing resources market, we take
the competition among miners~\cite{Kiayias2016} and network effects
of blockchain by nature~\cite{Catalini2016} into consideration.
We study the auction mechanism with allocative externalities\footnote{The allocative externalities occur when the allocation result of the
auction affects the valuation of the miners.} to maximize the social welfare. 
\item From the perspective of the CFP, we formulate social welfare maximization
problems for two bidding schemes: constant-demand scheme and multi-demand
scheme. For the constant-demand bidding scheme, we develop an optimal
algorithm that achieves optimal social welfare. For the multi-demand
bidding scheme, we prove that the formulated problem is NP-hard and
equivalent to the problem of non-monotone submodular maximization
with knapsack constraints. Therefore, we introduce an approximate
algorithm that generates sub-optimal social welfare. Both the algorithms
are designed to be truthful, individually rational and computationally
efficient. 
\item Based on the real-world mobile blockchain experiment, we define and
verify two characteristic functions for system model formulation.
One is the hash power function that describes the relationship between
the probability of successfully mining a block and the corresponding
miner's computing power. The other one is the network effects function
that characterizes the relationship between security of the blockchain
network and total computing resources invested into the network. 
\item Our simulation results show that the proposed auction mechanisms not
only help the CFP make practical and efficient computing resource
trading strategies, but also offer insightful guidance to the blockchain
developer in designing the blockchain protocol. 
\end{itemize}
To the best of our knowledge, this is the first work that investigates
resource management and pricing for blockchain networks in the auction-based
market. This paper is an extended version of our conference paper~\cite{Jiao2018}.
In\cite{Jiao2018}, \textcolor{black}{we considered only the} miners
with constant demand and did not perform the real-world experiment
to verify the network effects function.

The rest of this paper is organized as follows. Section~\ref{sec:Related-Work}
reviews related work. The system model of cloud/fog computing resource
market for blockchain networks is introduced in Section~\ref{sec:System-Model}.
Section~\ref{sec:SW_auction_unit_demand} discusses the constant-demand
bidding scheme and the optimal algorithm for social welfare maximization.
In Section~\ref{sec:Multi-bid auction}, the approximate algorithm
for multi-demand bidding scheme is presented in detail. Experimental
results of mobile blockchain and the performance analysis of the proposed
auction mechanisms are presented in Section~\ref{sec:Experiment-and-numerical}.
Finally, Section~\ref{sec:Conclusions} concludes the paper. Table
\ref{tab:notations} lists notations frequently used in the paper.
\begin{table}[tbh]
\caption{Frequently used notations.\label{tab:notations}}
\centering{}%
\begin{tabular}{|>{\raggedright}p{0.15\columnwidth}|>{\raggedright}p{0.7\columnwidth}|}
\hline 
\textbf{\noun{Notation}}  & \textbf{\noun{Description}}\tabularnewline
\hline 
\hline 
$\mathcal{N}$, $N$  & Set of miners and the total number of miners\tabularnewline
\hline 
$\mathcal{M}$  & Set of winners, i.e., the selected miners by the auction\tabularnewline
\hline 
$\mathbf{d}$, $d_{i}$  & Miners' service demand profile and miner $i$'s demand for cloud/fog
computing resource \tabularnewline
\hline 
$\mathbf{b}$, $b_{i}$  & Miners' bid profile and miner $i$'s bid for its demand $d_{i}$\tabularnewline
\hline 
$\mathbf{x}$, $x_{i}$  & Resource allocation profile and allocation result for miner $i$\tabularnewline
\hline 
$\mathbf{p}$, $p_{i}$  & Price profile and cloud/fog computing service price for miner $i$\tabularnewline
\hline 
$\gamma_{i}$  & Miner $i$'s hash power\tabularnewline
\hline 
$T$, $r$  & Fixed bonus from mining a new block and the transaction fee rate\tabularnewline
\hline 
$s_{i}$  & Miner $i$'s block size\tabularnewline
\hline 
\emph{$\lambda$}  & Average block time\tabularnewline
\hline 
$D$  & Total supply of computing resources from CFP\tabularnewline
\hline 
$w$  & Network effects function\tabularnewline
\hline 
$q$  & Quantity of computing resource required by constant-demand miner\tabularnewline
\hline 
$\beta$  & Demand constraint ratio for multi-demand miner\tabularnewline
\hline 
\end{tabular}
\end{table}

\section{Related Work\label{sec:Related-Work}}

As the core part of the blockchain network, creating blocks integrates
the distributed database (i.e., ledger), the consensus protocol and
the executable scripts (i.e., smart contract)~\cite{Anh2018}. From
the perspective of data processing, a DApp is essentially developed
on the basis of smart contracts and transactional data stored in the
blockchain. DApps usually use the distributed ledger to monitor the
state/ownership changes of the tokenized assets. The implementation
of smart contracts are driven by the transaction/data change to autonomously
determine the blockchain state transition, e.g., the asset re-distribution
among the DApp users~\cite{Christidis2016,Anh2018}. With the public
blockchain, DApps do not have to rely on a centralized infrastructure
and intermediary that supports ledger maintenance and smart contracts
execution with dedicated storage and computing resources. Instead,
DApp providers adopt the token-based reward mechanisms which incentivize
people to undertake the tasks of resource provision and system maintenance.
In this way, the functionalities of DApps can be freely activated
and realized among transaction issuing/validation, information propagation/storage
and consensus participation~\cite{Anh2018,Tschorsch2016}. Therefore,
the public blockchain network is a suitable platform for incentive-driven
Distributed Autonomous Organization (DAO) systems. To date, a line
of literature study the DAO in wireless networking based on the public
blockchain. The authors in~\cite{Chatzopoulos2017} established a
trading platform for Device-to-Device (D2D) computation offloading
based on a dedicated cryptocurrency network. They introduced smart
contract-based auctions between neighbor D2D nodes to execute resource
offloading and offload the block mining tasks to the cloudlets. The
authors in~\cite{Kopp2017} adopted a PoW-based public blockchain
as the backbone of a P2P file storage market, where the privacy of
different parties in a transaction is enhanced by the techniques such
as ring signatures and one-time payment addresses. When identity verification
is required for market access granting, e.g., in the scenarios of
autonomous network slice brokering~\cite{Backman2017} and P2P electricity
trading~\cite{Kang2017}, the public blockchain can be adapted into
consortium blockchain by introducing membership authorizing servers
with little modification to the consensus protocols and the smart
contract design.

Recently, there have already been some studies on the blockchain network
from the point of game theory. The authors in~\cite{Houy2016} proposed
a game-theoretic model where the occurrence of working out the PoW
puzzle was modeled as a Poisson process. Since a miner's expected
reward largely depends on the block size, each miner's response is
to choose a reasonable block size before mining for its optimal expected
reward. An analytical Nash equilibrium in a two-player case was discussed.
In~\cite{Lewenberg2015}, the authors presented a cooperative game
model to investigate the mining pool. In the pool, miners form a coalition
to accumulate their computing power for steady rewards. Nevertheless,
these works mainly focused on the block mining strategies and paid
little attention to the deployment of the blockchain network for developing
DApps and corresponding resource allocation problems. As a branch
of the game theory, the auction mechanism has been widely used to
deal with resource allocation issues in various areas, such as mobile
crowdsensing~\cite{Zhang2016,Yang2016,Jin2015}, cloud/edge computing~\cite{Mashayekhy2015,Kiani2017},
and spectrum trading~\cite{Zheng2015}. In~\cite{Jin2015}, the
authors proposed incentive mechanisms for efficient mobile task crowdsourcing
based on reverse combinatorial auctions. They considered data quality
constraints in a linear social welfare maximization problem. The authors
in~\cite{Mashayekhy2015} designed optimal and approximate strategy-proof
mechanisms to solve the problem of physical machine resource management
in clouds. They formulated the problem as a linear integer program.
In~\cite{Kiani2017}, the authors proposed an auction-based profit
maximization model for hierarchical mobile edge computing. Unfortunately,
it did not take any economic properties, e.g., incentive compatibility,
into account. While guaranteeing the strategyproofness, the authors
in~\cite{Zheng2015} investigated the problem of redistributing wireless
channels and focused on the social welfare maximization. They not
only considered strategyproofness, but also took the channel spatial
reusability, channel heterogeneity and bid diversity into account.
However, in their combinatorial auction setting, the bidder's requested
spectrum bundle is assumed to be always truthful. In fact, none of
these works can be directly applied to allocating computing resources
for the blockchain mainly due to its unique architecture. In the blockchain
network, the allocative externalities~\cite{Salek2008,Jehiel2001}
should be particularly taken into consideration. For example, besides
its own received computing resources, each miner also cares much about
the other miners' computing power.

In our paper, the social welfare optimization in the multi-demand
bidding scheme is proved to be a problem of non-monotone submodular
maximization with knapsack constraints, which has not been well studied
in auction mechanism design to date. The most closely related papers
are~\cite{Yang2016} and~\cite{Zhao2014} in mobile crowdsourcing.
In~\cite{Yang2016}, the authors presented a representative truthful
auction mechanism for crowdsourcing tasks. They studied a non-monotone
submodular maximization problem without constraints. In~\cite{Zhao2014},
the authors formulated a monotone sub-modular function maximization
problem when designing a truthful auction mechanism. The total payment
to the mobile users is constrained by a fixed budget. Technically,
the algorithms in aforementioned works cannot be applied in our models.
In addition, the authors in~\cite{Luong2018} used deep learning
to recover the classical optimal auction for revenue maximization
and applied it in the edge computing resources allocation in mobile
blockchain. However, it only considers one unit of resource in the
auction. 

\section{System Model: Blockchain Mining and Auction Based Market Model\label{sec:System-Model}}

\subsection{Cloud/Fog Computing Resource Trading \label{subsec:Edge-Computing-Resources}}

\textcolor{black}{Our system model is built under the assumptions
that 1) the public blockchain network adopts the classical PoW} consensus
protocol~\cite{Nakamoto2008}, 2) miners do not use their own devices,
e.g., computationally lightweight or mobile devices, to execute the
mining tasks. We consider a scenario where there is one CFP and a
community of miners $\mathcal{N=}\{1,\ldots,N\}$. Each miner runs
a blockchain-based DApps to record and verify the transactional data
sent to the blockchain network. Due to insufficient energy and computing
capacity of their devices, the miners offload the task of solving
PoW to nearby cloud/fog computing service which is deployed and maintained
by the CFP. To perform the trading, the CFP launches an auction. The
CFP first announces auction rules and the available service to miners.
Then, the miners submit their resource demand profile $\mathbf{d}=(d_{1},\ldots,d_{N})$
and corresponding bid profile $\mathbf{b}=(b_{1},\ldots,b_{N})$ which
represents the valuations of their requested resources. After having
received miners' demands and bids, the CFP selects the winning miners
and notifies all miners the allocation $\mathbf{x}=(x_{1},\ldots,x_{N})$
and the service price $\mathbf{p}=(p_{1},\ldots,p_{N})$, i.e., the
payment for each miner\footnote{Throughout this paper, the terms price and payment are used interchangeably.}.
We assume that miners are single minded~\cite{Nisan2007}, that is,
each miner only accepts its requested quantity of resources or none.
The setting $x_{i}=1$ means that miner $i$ is within the winner
list and allocated resources for which it submits the bid, while $x_{i}=0$
means no resource allocated. The payment for a miner which fails the
auction is set to be zero, i.e., $p_{i}=0$ if $x_{i}=0$. At the
end of the auction, the selected miners or winners make the payment
according to the price assigned by the CFP and access the cloud/fog
computing service.

\subsection{Blockchain Mining with Cloud/Fog Computing Service \label{subsec:Blockchain-Mining-with-ES} }

With the allocation $x_{i}$ and demand $d_{i}$, miner $i$'s hash
power $\gamma_{i}$ can be calculated from 
\begin{equation}
\gamma_{i}(\mathbf{d},\mathbf{x})=\frac{d_{i}x_{i}}{d_{\mathcal{N}}},\label{eq:hash-power}
\end{equation}
which is a linear fractional function. The function depends on other
miners' allocated computing resources and satisfies $\sum_{i\in\mathcal{N}}\gamma_{i}=1$.
$d_{\mathcal{N}}=\sum_{i\in\mathcal{N}}d_{i}x_{i}$ is the total quantity
of allocated resources. The hash power function $\gamma_{i}(\mathbf{d},\mathbf{x})$
is verified by a real-world experiment as presented later in Section~\ref{sec:Experiment-and-numerical}.

Before executing the miner selection by the auction, each miner has
collected unconfirmed transactional data into its own block. We denote
each miner's \emph{block size}, i.e., the total size of transactional
data and metadata, by $\mathbf{s}=(s_{1},\ldots,s_{N})$. In the mining
tournament, the generation of new blocks follows a Poisson process
with a constant mean rate $\frac{1}{\lambda}$ throughout the whole
blockchain network~\cite{Kraft2016}. $\lambda$ is also known as
the \emph{average} \emph{block time}. If the miner $i$ finds a new
block, the time for propagation and verification of transactions in
the block is dominantly affected by $s_{i}$. The first miner which
successfully has its block reach consensus can receive a \emph{token
reward} $R$. The token reward is composed of a \emph{fixed bonus}
$T\geq0$ for mining a new block and a variable transaction fee $t_{i}=rs_{i}$
determined by miner $i$'s block size $s_{i}$ and a predefined \emph{transaction
fee rate} $r$~\cite{Houy2016}. Thus, miner $i$'s token reward
$R_{i}$ can be expressed as follows: 
\begin{equation}
R_{i}=(T+rs_{i})\mathbb{P}_{i}(\gamma_{i}(\mathbf{d},\mathbf{x}),s_{i}),\label{eq:Expected-reward}
\end{equation}
where $\mathbb{P}_{i}(\gamma_{i}(\mathbf{d},\mathbf{x}),s_{i})$ is
the probability that miner $i$ receives the reward for contributing
a block to the blockchain.

We note that obtaining the reward rests with successful mining and
instant propagation. Miner $i$'s probability of discovering the nonce
value $P_{i}^{m}$ is equal to its hash power $\gamma_{i}$, i.e.,
$P_{i}^{m}=\gamma_{i}$. However, a lucky miner may even lose the
tournament if its broadcast block is not accepted by other miners
at once, i.e., failing to reach consensus. The newly mined block that
cannot be added onto the blockchain is called \emph{orphan block}~\cite{Houy2016}.
A larger block needs more propagation and verification time, thus
resulting in larger delay in reaching consensus. As such, a larger
block size means a higher chance that the block suffers orphaned.
According to the statistics displayed in~\cite{Narayanan2016}, miner
$i$'s block propagation time $\tau_{i}$ is linear to the block size,
i.e., $\tau_{i}=\xi s_{i}$. $\xi$ is a constant that reflects the
impact of $s_{i}$ on $\tau_{i}$. Since the arrival rate of new blocks
follows the Poisson distribution, miner $i$'s orphaning probability
is: 
\begin{equation}
P_{i}^{o}=1-\mathrm{e^{-\frac{1}{\lambda}\tau_{i}}}.\label{eq:p-orphan}
\end{equation}
Substituting $\tau_{i}$, we can express $\mathbb{P}_{i}$ as follows:
\begin{align}
\mathbb{P}_{i} & (\gamma_{i}(\mathbf{d},\mathbf{x}),s_{i})=P_{i}^{m}(1-P_{i}^{o})=\gamma_{i}\mathrm{e}^{-\frac{1}{\lambda}\xi s_{i}}.\label{eq:p-mining-a-block}
\end{align}

\subsection{Business Ecosystem for Blockchain based DApps}

\begin{figure}[tbh]
\begin{centering}
\includegraphics[width=1\columnwidth]{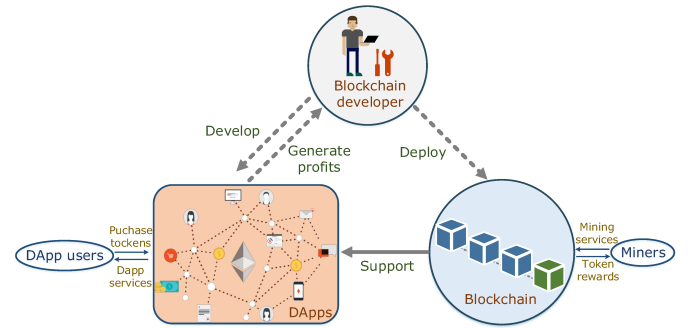} 
\par\end{centering}
\caption{Business ecosystem for blockchain based DApps.\label{fig:DApps}}
\end{figure}
Here, we describe the business ecosystem for blockchain based DApps
in Fig.~\ref{fig:DApps}. In developing a blockchain based DApps,
there exists a blockchain developer which is responsible for designing
or adopting the blockchain operation protocol. The developer specifies
the fixed bonus $T$, the transaction fee rate $r$ and so on. Through
adjusting the difficulty of finding the new nonce, the blockchain
developer keeps the average block time $\lambda$ at a reasonable
constant value. To support the DApps, in the deployed blockchain network,
miners perform mining and token reward, i.e., $R$, is used to incentivize
them. The reward may come from the token that DApps users pay to the
blockchain network.

When bidding for computing resources, miners always evaluate the value
of the tokens. In fact, the intrinsic value of tokens depends on the
trustworthiness and robustness, i.e., the value of the blockchain
network itself. From the perspective of trustworthiness, the PoW-based
blockchain is only as secure as the amount of computing power dedicated
to mining tasks~\cite{Catalini2016}. This results in positive network
effects~\cite{Catalini2016} in that as more miners participate and
more computing resources are invested, the security of the blockchain
network is improved, and hence the value of a reward given to miners
increases. A straightforward example is that if the robustness of
the blockchain network is very low, i.e., vulnerable to manipulation
(51\% attack, double-spending attack, etc.), that means this blockchain
is insecure and cannot support any decentralized application effectively.
Naturally, this blockchain network losses its value and its distributed
tokens (including the rewards to miners) would be worthless. On the
contrary, if there are many miners and computing resources invested,
the blockchain would be more reliable and secure~\cite{Aitzhan2016}.
Thus, users would trust it more and like to use its supported decentralized
applications through purchasing the tokens and then miners would also
gain more valuation on their received tokens (reward). To confirm
this fact, we conduct a real-world experiment (see Section~\ref{subsec:Verification-for-Hash-Net})
to evaluate the value of the tokens and the reward by examining the
impact of the total computing power on preventing double-spending
attacks. By curve fitting of the experimental data, we define the
network effects by a non-negative utility function as follows: 
\begin{equation}
w(\pi)=a_{1}\pi-a_{2}\pi\mathrm{e}^{a_{3}\pi},\label{eq:network_effects_function}
\end{equation}
where $\pi=\frac{d_{\mathcal{N}}}{D}\in[0,1]$ is the normalized total
computing power of the blockchain network. $d_{\mathcal{N}}=\sum_{i\in\mathcal{N}}d_{i}x_{i}$
is the total quantity of allocated computing resources, and $D$ is
the maximum quantity that CFP can supply. $a_{1},a_{2},a_{3}>0$ are
curve fitting parameters and this network effects function in the
feasible domain is monotonically increasing with a diminishing return.

\subsection{Miner's Valuation on Cloud/Fog Computing Resources}

\begin{figure*}[t]
\begin{centering}
\includegraphics[width=1.8\columnwidth]{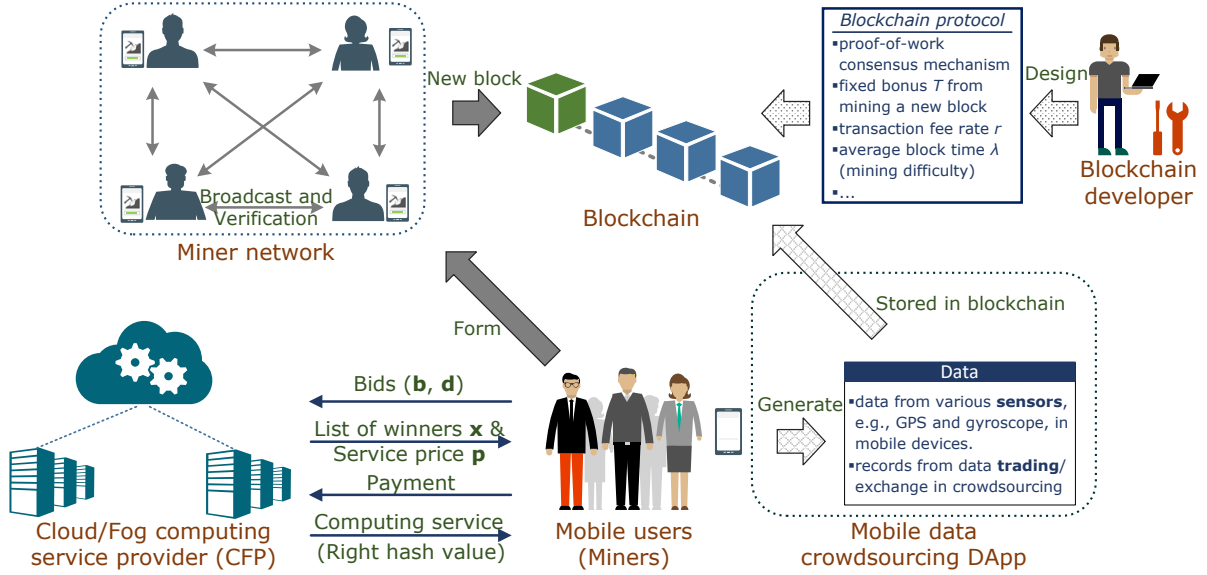} 
\par\end{centering}
\caption{An example mobile data crowdsourcing application illustrating the
system model and the cloud/fog computing resource market for blockchain
networks.\label{fig:running_example}}
\end{figure*}
In the auction, a miner's bid represents the valuation of computing
resources for which it demands. Since miner $i$ cannot know the number
of winning miners and the total quantity of allocated resources until
the end of auction, we assume that miner $i$ can only give the bid
$b_{i}$ according to its expected reward $R_{i}$ and demand $d_{i}$
without considering network effects and other miners' demands, i.e.,
setting $w(d_{\mathcal{N}})=1$ and $\sum_{j\in\mathcal{N}\setminus\{i\}}d_{j}x_{j}=0$.
In other words, miner $i$ has an \emph{ex-ante} valuation $v'_{i}$
which can be written as ($P_{i}^{m}=\gamma_{i}=1$): 
\begin{equation}
v'_{i}=R_{i}d_{i}=\left(T+rs_{i}\right)\mathrm{e}^{-\frac{1}{\lambda}\xi s_{i}}d_{i}.\label{eq:ex-ante}
\end{equation}
Here, we assume that $R_{i}$ represents the miner $i$'s valuation
for one unit computing resource and $d_{i}$ is decided according
to miner $i$'s own available budget. Since our proposed auction mechanisms
are truthful (to be proved later), $b_{i}$ is equal to the true \emph{ex-ante}
valuation $v'_{i}$, i.e., $b_{i}=v'_{i}$.

After the auction is completed, miners receive the allocation result,
i.e., $\mathbf{x}$, and are able to evaluate the network effects.
Hereby, miner $i$ has an {\emph{ex-post}} valuation $v''_{i}$
as follows: 
\begin{eqnarray}
v''_{i} & = & v'_{i}w(\pi)\gamma_{i}(\mathbf{d},\mathbf{x})\nonumber \\
 & = & \frac{d_{i}^{2}x_{i}}{d_{\mathcal{N}}}\left(a_{1}\pi-a_{2}\pi\mathrm{e}^{a_{3}\pi}\right)\left(T+rs_{i}\right)\mathrm{e}^{-\frac{1}{\lambda}\xi s_{i}}\nonumber \\
 & = & \frac{d_{i}^{2}x_{i}}{D}\left(a_{1}-a_{2}\mathrm{e}^{a_{3}\frac{d_{\mathcal{N}}}{D}}\right)\left(T+rs_{i}\right)\mathrm{e}^{-\frac{1}{\lambda}\xi s_{i}}.\label{eq:ex-post}
\end{eqnarray}

\subsection{Social Welfare Maximization\label{subsec:Social-Welfare-Maximization}}

The CFP selects winning miners, i.e., winners, and determines corresponding
prices in order to maximize the social welfare. Let $c$ denote the
unit cost of running the cloud/fog computing service, so the total
cost to the CFP can be expressed by $C(d_{\mathcal{N}})=cd_{\mathcal{N}}=\sum_{i\in\mathcal{N}}cd_{i}x_{i}$.
Thus, we define the social welfare of the blockchain network $S$
as the difference between the sum of all miners' ex-post valuations
and the CFP's total cost, i.e., 
\begin{eqnarray}
S(\mathbf{x}) & = & \sum_{i\in\mathcal{N}}v''_{i}-C(d_{\mathcal{N}})\nonumber \\
 & = & \sum_{i\in\mathcal{N}}\frac{d_{i}^{2}x_{i}}{D}\left(a_{1}-a_{2}\mathrm{e}^{a_{3}\frac{d_{\mathcal{N}}}{D}}\right)\left(T+rs_{i}\right)\mathrm{e}^{-\frac{1}{\lambda}\xi s_{i}}\nonumber \\
 &  & \qquad-cd_{\mathcal{N}}.\label{eq:def_social_welfare}
\end{eqnarray}
Therefore, the primary objective of designing the auction mechanism
is to solve the following integer programming:

\begin{eqnarray}
\max_{\mathbf{x}} &  & S(\mathbf{x})=\sum_{i\in\mathcal{N}}\left(\frac{d_{i}^{2}x_{i}}{D}\left(a_{1}-a_{2}\mathrm{e}^{\frac{a_{3}}{D}\sum_{i\in\mathcal{N}}d_{i}x_{i}}\right)\right.\nonumber \\
 &  & \qquad\qquad\left.\left(T+rs_{i}\right)\mathrm{e}^{-\frac{1}{\lambda}\xi s_{i}}\right)-\sum_{i\in\mathcal{N}}cd_{i}x_{i},\label{eq:socail-welfare-original}\\
\mathrm{s.t.} &  & \sum_{i\in\mathcal{N}}d_{i}x_{i}\leq D,\label{eq:constraint-supply}\\
 &  & x_{i}\in\{0,1\},\forall i\in\mathcal{N},\label{eq:binary_limit}
\end{eqnarray}
where (\ref{eq:constraint-supply}) is the constraint on the quantity
of computing resources that CFP can offer. In the next two sections,
we consider two types of bidding scheme in the auction design: constant-demand
bidding scheme and multi-demand bidding scheme. Accordingly, there
are two types of miners: constant-demand miners and multi-demand miners.
We aim to maximize the social welfare, while guaranteeing the truthfulness,
individual rationality and computational efficiency.

\subsection{Example Application: Mobile Data Crowdsourcing }

As shown in Fig. \ref{fig:running_example}, we take an example of
mobile data crowdsourcing to illustrate the use of our model and and
to demonstrate an effectiveness of the related concepts. Initially,
there are a group of mobile users. Each of the mobile users can be
either a worker that collects data from the sensors in its mobile
device or a requester that wants to buy the sensing data from other
users (workers). However, there is often no trusted or authorized
crowdsourcing platform to process the data trading and record the
transactions. Moreover, no mobile user has enough trust, right, or
capability to establish and operate such a centralized platform. In
this case, a viable solution is to design and deploy a blockchain
based crowdsourcing DApp by a blockchain developer. Based on the designed
protocol, the mobile users can utilize the available cloud/fog computing
resources to self-organize a reliable blockchain network. Thus, their
data trading activities can be facilitated by the established decentralized
crowdsourcing platform with smart contracts.

The blockchain developer adopts the PoW protocol and sets the parameters,
such as the fixed reward $T$, the transaction fee rate $r$ and the
average block time $\lambda$. Due to limited energy and computational
capability, mobile users (miners) need to buy computing resources
from the CFP through an auction process and then join the miner network.
Before the auction begins, miner $i$ may possess a certain amount
of data to be stored in the blockchain and knows its block size $s_{i}$.
According to (\ref{eq:ex-ante}), the miner $i$ will evaluate its
expected reward and the ex-ante value $v'_{i}$ of the computing resources
based on the protocol parameters, its block size and demand. Next,
the miner $i$ submits the bid $b_{i}$ and the demand $d_{i}$ to
the CFP. Using our proposed auction algorithm, the CFP can select
the winning miners, i.e., the allocation $x_{i}$, and determine the
price $p_{i}$ to maximize the social welfare. Meanwhile, it can guarantee
the miner's truthfulness and non-negative utility which is the difference
between the ex-post valuation $v''_{i}$ and its payment $p_{i}$.
Once the auction ends, the winning miners which are allocated the
computing resources form a miner network. With the CFP service in
solving the PoW puzzle and calculating the hash values, the winning
miners can start the mining and consensus process to verify and contribute
new blocks containing the crowdsourced data and corresponding transaction
records to the blockchain. For more details about the blockchain-based
crowdsourcing, please refer to~\cite{Li2018}.

\section{Auction-based Mechanism for Constant-demand Miners\label{sec:SW_auction_unit_demand}}

In this section, we first consider a simple case where all miners
submit bids for the same quantity of computing resources. Here, each
miner's demand is $q$ units, i.e., $d_{i}=q\in(0,D),\forall i\in\mathcal{N}$.
Thus, the optimization problem for the CFP can be expressed as follows:

\begin{eqnarray}
\max_{\mathbf{x}} &  & S(\mathbf{x})=\sum_{i\in\mathcal{N}}\left(\frac{q^{2}x_{i}}{D}\left(a_{1}-a_{2}\mathrm{e}^{\frac{a_{3}}{D}\sum_{i\in\mathcal{N}}qx_{i}}\right)\right.\nonumber \\
 &  & \left.\qquad\qquad\left(T+rs_{i}\right)\mathrm{e}^{-\frac{1}{\lambda}\xi s_{i}}\right)-\sum_{i\in\mathcal{N}}cqx_{i},\label{eq:socail-welfare-original-unit}\\
\mathrm{s.t.} &  & \sum_{i\in\mathcal{N}}qx_{i}\leq D,\label{eq:constraint_supply_unit_demand}\\
 &  & x_{i}\in\{0,1\},\forall i\in\mathcal{N}.\label{eq:xi_integer_unit_demand}
\end{eqnarray}

\begin{algorithm}[tbh]
{\scriptsize{}\scriptsize \begin{algorithmic}[1] \Require{Miners' bid profile~$\mathbf{b}$ and demand profile~$\mathbf{d}$.} \Ensure{Resource allocation $\mathbf{x}$ and service price~$\mathbf{p}$.} \Begin 	\ForEach{$i \in \mathcal{N}$} 		\State{$x_i \gets 0$, $p_i \gets 0$} 	\EndFor 	\State{Sort bids $\mathbf{b}$ in descending order.} 	\State{$j \gets {\arg\max}_{j \in \mathcal{N}}b_{j}$} 	\State{$\mathcal{M} \gets \{j\}$, $S \gets \frac{q}{D}\left(a_{1}-a_{2}\mathrm{e}^{\frac{a_{3}q}{D}}\right)b_{j}-cq$} 	\While{$\mathcal{M}\neq\mathcal{N}$ \textbf{and} ${\left|\mathcal{M}\right|} \leq D$} 		\State{$j \gets {\arg\max}_{j \in \mathcal{N}\setminus\mathcal{M}}b_{j}$} 		\State{$\mathcal{M}_{t} \gets \mathcal{M} \cup \{j\}$} 		\State{$S_t \gets \sum_{i\in\mathcal{M}_{t}}\frac{q}{D}\left(a_{1}-a_{2}\mathrm{e}^{a_{3}q\left|\mathcal{M}_{t}\right|}\right)b_{i}-cq\left|\mathcal{M}_{t}\right|$} 		\If{$S_t < S$ \textbf{or} $S_t < 0$} 			\Break 		\EndIf 		\State{$\mathcal{M} \gets \mathcal{M} \cup \{j\}$} 	\EndWhile
	\ForEach{$i \in \mathcal{M}$} 		\State{$x_i \gets 1$, $\mathcal{N}_{-i} \gets \mathcal{N} \setminus \{i\}$, $\mathcal{M}_{-i} \gets \mathcal{M} \setminus \{i\}$} 		\State{$j \gets {\arg\max}_{j \in \mathcal{N}_{-i}}b_{j}$} 		\State{$\mathcal{M}' \gets \{j\}$, $S' \gets \frac{q}{D}\left(a_{1}-a_{2}\mathrm{e}^{\frac{a_{3}q\left|\mathcal{M}'\right|}{D}}\right)b_{j}-cq$}
		\While{$\mathcal{M}'\neq\mathcal{N}$ \textbf{and} ${\left|\mathcal{M}'\right|} \leq D$} 		\State{$j \gets {\arg\max}_{i \in \mathcal{N}_{-i}\setminus\mathcal{M}'}b_{j}$} 		\State{$\mathcal{M}'_{t} \gets \mathcal{M}' \cup \{j\}$} 		\State{$S'_t \gets \sum_{i\in\mathcal{M}'_{t}}\frac{q}{D}\left(a_{1}-a_{2}\mathrm{e}^{\frac{a_{3}q\left|\mathcal{M}'_{t}\right|}{D}}\right)b_{i}-cq\left|\mathcal{M}'_{t}\right|$} 		\If{$S'_t < S'$ \textbf{or} $S'_t < 0$} 			\Break 		\EndIf 		\State{$\mathcal{M}' \gets \mathcal{M}'_{t}$, $S' \gets S'_{t}$} 	\EndWhile \State{$p_i=S'-\sum_{i\in\mathcal{M}_{-i}}\frac{q}{D}\left(a_{1}-a_{2}\mathrm{e}^{\frac{a_{3}q\left|\mathcal{M}_{-i}\right|}{D}}\right)b_{i}-cq\left|\mathcal{M}_{-i}\right|$} 	\EndFor \End \end{algorithmic}\caption{CDB auction\label{alg:1}}
}
\end{algorithm}
The first proposed truthful auction for Constant-Demand miners in
Blockchain networks (CDB auction), as presented in Algorithm~\ref{alg:1},
is an optimal one and its rationale is based on the well-known Myerson's
characterization~\cite{Myerson1981} provided in Theorem~\ref{thm:unit-truthful-condition}. 
\begin{thm}
(\cite[Theorem~13.6]{Nisan2007}) An auction mechanism is truthful
if and only if it satisfies the following two properties:\label{thm:unit-truthful-condition} 
\end{thm}
\begin{enumerate}
\item \emph{Monotonicity: If miner $i$ wins the auction with bid $b_{i}$,
then it will also win with any higher bid $b_{i}'>b_{i}$.} 
\item \emph{Critical payment: The payment by a winner is the smallest value
needed in order to win the auction. } 
\end{enumerate}
As illustrated in Algorithm~\ref{alg:1}, the CDB auction consists
of two consecutive processes: winner selection (lines~5-16) and service
price calculation (lines~17-31). The winner selection process is
implemented with a greedy method. For the convenience of later discussion,
we define a set of winners as $\mathcal{M}$. Adding a miner $i$
in $\mathcal{M}$ means setting $x_{i}=1$. Thus, we transform the
original problem in (\ref{eq:socail-welfare-original-unit})-(\ref{eq:xi_integer_unit_demand})
to an equivalent set function form as follows: 
\begin{eqnarray}
\max_{\mathcal{M\subseteq N}} &  & S(\mathcal{M})=\sum_{i\in\mathcal{M}}\left(a_{1}-a_{2}\mathrm{e}^{\frac{a_{3}q\left|\mathcal{M}\right|}{D}}\right)\frac{qb_{i}}{D}-cq\left|\mathcal{M}\right|,\nonumber \\
\label{eq:equivalent_unit_demand}\\
\mathrm{s.t.} &  & q\mathcal{\left|M\right|}\leq D,\label{eq:equivalent_constraint}
\end{eqnarray}
where $\left|\mathcal{M}\right|$ represents the cardinality of set
$\mathcal{M}$ which is the number of winners in $\mathcal{M}$ and
$b_{i}=v'_{i}=\left(T+rs_{i}\right)\mathrm{e}^{-\frac{1}{\lambda}\xi s_{i}}q$.
In the winner selection process (lines~5-11), miners are first sorted
in a descending order according to their bids. Then, they are sequentially
added to the set of winners $\mathcal{M}$ until the social welfare
$S(\mathcal{M})$ begins to decrease. Finally, the set of winners
$\mathcal{M}$ and the allocation $\mathbf{x}$ are output by the
algorithm. 
\begin{prop}
The resource allocation $\mathbf{x}$ output by Algorithm~\ref{alg:1}
is globally optimal to the social welfare maximization problem given
in (\ref{eq:socail-welfare-original-unit})-(\ref{eq:xi_integer_unit_demand}).\label{prop:unit_demand_optimal} 
\end{prop}
\begin{IEEEproof}
With the proof by contradiction, this result follows from Claim~\ref{claim:3}. 
\end{IEEEproof}
\begin{claim}
Let $\mathcal{M}_{A}$ be the solution output by Algorithm~\ref{alg:1}
on input $\mathbf{b}$, and $\mathcal{M}_{O}$ be the optimal solution.
If $\mathcal{M}_{A}\neq\mathcal{M}_{O}$, then we can construct another
solution $\mathcal{M}_{O}^{*}$ whose social welfare $S(\mathcal{M}_{O}^{*})$
is even larger than the optimal social welfare $S(\mathcal{M}_{O})$.\label{claim:3} 
\end{claim}
\begin{IEEEproof}
We assume $b_{1}\geq\cdots\geq b_{N}$ and $\mathcal{M}_{A}\neq\mathcal{M}_{O}$.
Next, we consider two cases.

1) Case 1: $\mathcal{M}_{O}\subset\mathcal{M}_{A}$. According to
Algorithm~\ref{alg:1}, it is obvious that we can construct a solution
$\mathcal{M}_{O}^{*}$ with higher social welfare by adding a member
from $\mathcal{M}_{A}$ to $\mathcal{M}_{O}$.

2) Case 2: $\mathcal{M}_{O}\not\subset\mathcal{M}_{A}$. Let $m$
be the first element (while-loop lines~7-14) that $m\notin\mathcal{M}_{O}$.
Since $m$ is maximal ($b_{m}$ is minimal by assumption), we have
$1,\ldots,m-1\in\mathcal{M}_{O}$ and the corresponding set of winning
bids $\mathbf{b}_{\mathcal{M}_{O}}=\{b_{1},\ldots,b_{m-1},b'_{m},b'_{m+1},\ldots,b'_{\left|\mathcal{M}_{O}\right|}\}$,
where the bids $\{b_{1},\ldots,b'_{\left|\mathcal{M}_{O}\right|}\}$
are listed in the descending order. Meanwhile, Algorithm~\ref{alg:1}
chooses $\mathbf{b}_{\mathcal{W}_{A}}=\{b_{1},\ldots,b_{m-1},b{}_{m},b{}_{m+1},\ldots,b{}_{\left|\mathcal{M}_{A}\right|}\}$
and there must be $b_{m}>b'_{j}$ for all $j\geq m$. In particular,
we have $b_{m}>b'_{m}$. Hence, we define $\mathbf{b}_{\mathcal{M}_{O}^{*}}=\mathbf{b}_{\mathcal{M}_{O}}\cup\{b_{m}\}\setminus\{b'_{m}\}$
, i.e., we obtain $\mathbf{b}_{\mathcal{M}_{O}^{*}}$ by removing
$b'_{m}$ and adding $b_{m}$ to $\mathbf{b}_{\mathcal{M}_{O}}$.
Thus, the social welfare of $\mathbf{b}_{\mathcal{W}_{O}^{*}}$ is
calculated as follows: 
\[
S(\mathcal{M}_{O}^{*})=S(\mathcal{M}_{O})+\frac{q}{D}\left(a_{1}-a_{2}\mathrm{e}^{\frac{a_{3}q\left|\mathcal{M}\right|}{D}}\right)(b_{m}-b'_{m}).
\]
As $b_{m}-b'_{m}>0$, $(a_{1}-a_{2}\mathrm{e}^{\frac{a_{3}q\left|\mathcal{M}\right|}{D}})\frac{q}{D}>0$
and $\left|\mathcal{M}_{O}^{*}\right|=\left|\mathcal{M}_{O}\right|$,
$S(\mathcal{M}_{O}^{*})$ is strictly larger than $S(\mathcal{M}_{O})$.
This is in contradiction to that $\mathcal{M}_{O}$ is the optimal
solution and thus proves the claim. 
\end{IEEEproof}
We apply Vickrey--Clarke--Groves (VCG) mechanism~\cite{Krishna2009}
in the service price calculation. In lines~16-30, for each iteration,
we exclude one selected miner from the set of winners and re-execute
the winner selection process to calculate the social cost of the miner
as its payment. The VCG-based payment function is defined as follows:
\begin{equation}
p_{i}=S(\mathcal{M}_{\mathcal{N}\setminus\{i\}})-S(\mathcal{M}_{\mathcal{N}}\setminus\{i\}),\label{eq:VCG_payment}
\end{equation}
where $S(\mathcal{M}_{\mathcal{N}\setminus\{i\}})$ is the optimal
social welfare obtained when the selected miner $i$ is excluded from
the miner set $\mathcal{N}$, and $S(\mathcal{M}_{\mathcal{N}}\setminus\{i\})$
is the social welfare of the set of winners which is obtained by removing
miner $i$ from the optimal winner set selected from $\mathcal{N}$.
\begin{prop}
The CDB auction (Algorithm~\ref{alg:1}) is truthful. 
\end{prop}
\begin{IEEEproof}
Since the payment calculation in the algorithm relies on the VCG mechanism,
it directly satisfies the second condition in Theorem~\ref{thm:unit-truthful-condition}~\cite{Nisan2007}.
For the first condition about monotonicity in Theorem~\ref{thm:unit-truthful-condition},
we need to show that if a winning miner $i$ raises its bid from $b_{i}$
to $b_{i}^{+}$ where $b_{i}^{+}>b_{i}$, it still stays in the winner
set. We denote the original winner set by $\mathcal{M}$ and the new
winner set by $\mathcal{M}_{+}$ after miner $i$ changes its bid
to $b_{i}^{+}$. The original set of bids is $\mathbf{b}=\{b_{1},\ldots,b_{i},\ldots,b_{N}\}$
$(i\leq\mathcal{\left|M\right|})$ sorted in the descending order.
In addition, we define $S(\mathbf{b}_{\mathcal{K}})=S(\mathcal{K}),\forall\mathcal{K}\subseteq\mathcal{N}$
which means the social welfare of a set of bids is equal to that of
the set of corresponding miners. We discuss the monotonicity in two
cases.

1) Case 1: $b_{i-1}\geq b_{i}^{+}\geq b_{i}\geq b_{i+1}$. The new
set of ordered bids is $\mathbf{b}^{+}=\{b_{1},\ldots,b_{i-1},b_{i}^{+},b_{i+1},\ldots,b_{N}\}$.
We have 
\begin{align}
S(\{b_{1},\ldots,b_{i}^{+}\})=\frac{q}{D}\left(a_{1}-a_{2}\mathrm{e}^{\frac{a_{3}qi}{D}}\right)\left(\sum_{j=1}^{i-1}b_{j}+b_{i}^{+}\right)-cqi\nonumber \\
>S(\{b_{1},\ldots,b_{i}\})=\frac{q}{D}\left(a_{1}-a_{2}\mathrm{e}^{\frac{a_{3}qi}{D}}\right)\sum_{j=1}^{i}b_{j}-cqi.\label{eq:S-case1}
\end{align}
The social welfare of the new set of bids $\{b_{1},\ldots,b_{i}^{+}\}$
is larger than that of the original set of bids $\{b_{1},\ldots,b_{i}\}$,
which guarantees $b_{i}^{+}$ being in the set of winning bids.

2) Case 2: $b_{k-1}\geq b_{i}^{+}\geq b_{k}\geq\cdots\geq b_{i}$,
$1<k<i$. The new set of ordered bids is $\mathbf{b}^{+}=\{b_{1},\ldots,b_{k-1},b_{i}^{+},b_{k},\ldots,b_{i+1},\ldots,b_{N}\}$.
We have 
\begin{align}
S(\{b_{1},\ldots,b_{k-1},b_{i}^{+}\})= & \frac{q}{D}\left(a_{1}-a_{2}\mathrm{e}^{\frac{a_{3}qk}{D}}\right)\left(\sum_{j=1}^{k-1}b_{j}+b_{i}^{+}\right)\nonumber \\
 & \qquad-cqk,\label{eq:S-case2-k}
\end{align}

\begin{equation}
S(\{b_{1},\ldots,b_{k-1},b_{k}\})=\frac{q}{D}\left(a_{1}-a_{2}\mathrm{e}^{\frac{a_{3}qk}{D}}\right)\sum_{j=1}^{k}b_{j}-cqk,\label{eq:S-case-original-k}
\end{equation}
\begin{equation}
S(\{b_{1},\ldots,b_{k-1}\})=\frac{q}{D}\left(a_{1}-a_{2}\mathrm{e}^{\frac{a_{3}q(k-1)}{D}}\right)\sum_{j=1}^{k-1}b_{j}-cq(k-1).\label{eq:S-case2-k-1}
\end{equation}

As the coefficient $\frac{q}{D}\left(a_{1}-a_{2}\mathrm{e}^{\frac{a_{3}q\left|\mathcal{M}\right|}{D}}\right)$
in $S(\mathcal{M})$ is a monotonically decreasing function of $\mathcal{M}$,
increasing $b_{i}$ may change the set of winners $\mathcal{M}$ and
reduce the number of winning miners. However, the first $i$ bids
$\{b_{1},\ldots,b_{k-1},b_{k},\ldots,b_{i}\}$ in the original set
of bids $\mathbf{b}$ have already won the auction, so we have $S(\{b_{1},\ldots,b_{k-1},b_{k}\})>S(\{b_{1},\ldots,b_{k-1}\})$.
From the following inequation~(\ref{eq:inequality}), 
\begin{align}
S(\{b_{1},\ldots,b_{k-1},b_{k}\})=\frac{q}{D}\left(a_{1}-a_{2}\mathrm{e}^{\frac{a_{3}qk}{D}}\right)\left(\sum_{j=1}^{k-1}b_{j}+b_{k}\right)\nonumber \\
<\frac{q}{D}\left(a_{1}-a_{2}\mathrm{e}^{\frac{a_{3}qk}{D}}\right)\left(\sum_{j=1}^{k-1}b_{j}+b_{i}^{+}\right)=S\left(\{b_{1},\ldots,b_{k-1},b_{i}^{+}\}\right)\label{eq:inequality}
\end{align}
the proof can be finally concluded by 
\begin{equation}
S(\{b_{1},\ldots,b_{k-1},b_{i}^{+}\})>S(\{b_{1},\ldots,b_{k-1}\}),\label{eq:udmb_truthful_nouse}
\end{equation}
which implies that $b_{i}^{+}$ still remains the bid of a winner
in the auction. 
\end{IEEEproof}
\begin{prop}
The CDB auction (Algorithm~\ref{alg:1}) is computationally efficient
and individually rational. 
\end{prop}
\begin{IEEEproof}
Sorting the bids has the complexity of $O(N\log N)$. Since the number
of winners is at most $\min(\frac{D}{q},N)$, the time complexity
of the winner selection process (while-loop, lines~7-15) is $O(\min^{2}(\frac{D}{q},N))$.
In each iteration of the payment calculation process (lines~16-30),
a similar winner selection process is executed. Therefore, the whole
auction process can be performed in polynomial time with the time
complexity of $O(\min^{3}(\frac{D}{q},N)+N\log N)$.

According to Proposition~\ref{prop:unit_demand_optimal} and the
properties of the VCG mechanism~\cite{Krishna2009}, the payment
scheme in Algorithm~\ref{alg:1} guarantees the individual rationality. 
\end{IEEEproof}

\section{Auction-based Mechanisms for Multi-demand Miners\label{sec:Multi-bid auction}}

In this section, we investigate a more general scenario where miners
request multiple demands of cloud/fog computing resources.

\subsection{Social Welfare Maximization for the Blockchain Network}

We first investigate the winner selection problem defined in~(\ref{eq:socail-welfare-original})-(\ref{eq:binary_limit})
from the perspective of an optimization problem. Evidently, it is
a nonlinear integer programming problem with linear constraints, which
is NP-hard to obtain the optimal solution. Naturally, we can find
an approximate method with a lower bound guarantee. Similar to Section~\ref{sec:SW_auction_unit_demand},
the original problem is rewritten as a subset function form: 
\begin{eqnarray}
\max_{\mathcal{\mathcal{M}\subseteq N}} &  & S(\mathcal{M})=\sum_{i\in\mathcal{M}}\frac{d_{i}}{D}\left(a_{1}-a_{2}\mathrm{e}^{\frac{a_{3}\sum_{i\in\mathcal{M}}d_{i}}{D}}\right)b_{i}\nonumber \\
 &  & \qquad\qquad-c\sum_{i\in\mathcal{M}}d_{i},\label{eq:equivalent_multi_demand}\\
s.t. &  & \sum_{i\in\mathcal{M}}d_{i}\leq D,\label{eq:constraint_multi_demand}
\end{eqnarray}
where $S(\mathcal{M})$ is the social welfare function of the selected
set of winners $\mathcal{M}$ and $b_{i}=v'_{i}=\left(T+rs_{i}\right)\mathrm{e}^{-\frac{1}{\lambda}\xi s_{i}}d_{i}$.
This form means that we can view it as a subset sum problem~\cite{Lagarias1985}.
We assume that there is at least one miner $i$ such that $S(\{i\})>0$.
Additionally, although the miners can submit demands that they want
instead of the same constant quantity of computing resources, it is
reasonable to assume that the CFP puts a restriction on the purchase
quantity, i.e., $\beta_{1}D<d_{i}\leq\beta_{2}D$, where $\beta_{1}D$,
$\beta_{2}D$ are respectively the lower and upper limit on each miner's
demand, and $0<\beta_{1}<\beta_{2}<1$ are predetermined demand constraint
ratios. Clearly, $S(\emptyset)=0$. 
\begin{defn}
\emph{\label{def:Submodular-Function}(Submodular Function}~\cite{Lovasz1983}\emph{).
Let $\mathcal{X}$ be a finite set. A function $f$ : $2^{\mathcal{X}}\rightarrow\mathbb{R}$
is submodular if 
\begin{equation}
f(\mathcal{A}\cup\left\{ x\right\} )-f(\mathcal{A})\geq f(\mathcal{B}\cup\left\{ x\right\} )-f(\mathcal{B}),\label{eq:def_submodular_1}
\end{equation}
for any $\mathcal{A}\subseteq\mathcal{B}\subseteq\mathcal{X}$ and
$x\in\mathcal{X}\setminus\mathcal{B}$, where $\mathbb{R}$ is the
set of reals. A useful equivalent definition is that $f$ is submodular
if and only if the derived set-function } 
\begin{equation}
f_{x}(\mathcal{A})=f(\mathcal{A}\cup\left\{ x\right\} )-f(\mathcal{A})\qquad(\mathcal{A}\subseteq\mathcal{X}\setminus\left\{ x\right\} )\label{eq:def_submodular_2}
\end{equation}
is monotonically decreasing for all $x\in\mathcal{X}$. 
\end{defn}
\begin{prop}
The social welfare function \textup{$S(\mathcal{M})$ in~(\ref{eq:equivalent_multi_demand})
is submodular.} 
\end{prop}
\begin{IEEEproof}
By Definition~\ref{def:Submodular-Function}, we need to show that
$S_{u}(\mathcal{M})$ in~(\ref{eq:S_u(mu)}) is monotonically decreasing,
for every $\mathcal{M}\subseteq\mathcal{N}$ and $u\in\mathcal{N}\setminus\mathcal{M}$.
\begin{figure*}
\centering 
\begin{align}
S_{u}(\mathcal{M}) & =S(\mathcal{M}\cup\left\{ u\right\} )-S(\mathcal{M})\label{eq:S_u(mu)_1}\\
 & =\sum_{i\in\mathcal{M}\cup\left\{ u\right\} }\frac{d_{i}}{D}\left(a_{1}-a_{2}\mathrm{e}^{\frac{a_{3}\sum_{i\in\mathcal{M}\cup\left\{ u\right\} }d_{i}}{D}}\right)b_{i}-\sum_{i\in\mathcal{M}}\frac{d_{i}}{D}\left(a_{1}-a_{2}\mathrm{e}^{\frac{a_{3}\sum_{i\in\mathcal{M}}d_{i}}{D}}\right)b_{i}-cd_{u}\label{eq:S_u(mu)_2}\\
 & =\underset{\text{\ding{192}}}{\underbrace{\left(\left(a_{1}-a_{2}\mathrm{e}^{\frac{a_{3}\sum_{i\in\mathcal{M}\cup\left\{ u\right\} }d_{i}}{D}}\right)-\left(a_{1}-a_{2}\mathrm{e}^{\frac{a_{3}\sum_{i\in\mathcal{M}}d_{i}}{D}}\right)\right)\sum_{i\in\mathcal{M}}\frac{d_{i}b_{i}}{D}}}\quad+\underset{\text{\ding{193}}}{\underbrace{\left(a_{1}-a_{2}\mathrm{e}^{\frac{a_{3}\sum_{i\in\mathcal{M}\cup\left\{ u\right\} }d_{i}}{D}}\right)\frac{d_{u}b_{u}}{D}-cd_{u}}}\label{eq:S_u(mu)}
\end{align}
\end{figure*}
Let $g(z)=a_{1}-a_{2}\mathrm{e}^{\frac{a_{3}}{D}z}$, where $z\in\mathbb{R}^{+}$.
Then, the first derivative and second derivative of $g(z)$ are expressed
respectively as follows: 
\begin{equation}
\frac{\mathrm{d}g(z)}{\mathrm{d}z}=-\frac{a_{2}a_{3}}{D}\mathrm{e}^{\frac{a_{3}}{D}z},\frac{\mathrm{d^{2}}g(z)}{\mathrm{d}z^{2}}=-\frac{a_{2}a_{3}^{2}}{D^{2}}\mathrm{e}^{\frac{a_{3}}{D}z}.\label{eq:1_st_and_2_nd_derivative}
\end{equation}
Because $a_{2},a_{3},D>0$, we have $-\frac{a_{2}a_{3}}{D}\mathrm{e}^{\frac{a_{3}}{D}z}<0$
and $-\frac{a_{2}a_{3}^{2}}{D^{2}}\mathrm{e}^{\frac{a_{3}}{D}z}<0$,
which indicates that $g(z)$ is monotonically decreasing and concave.

Next, we discuss the monotonicity of $S_{u}(\mathcal{M})$ in~(\ref{eq:S_u(mu)}).
Note that expanding $\mathcal{M}$ means increasing the total quantity
of allocated resources $d_{\mathcal{M}}=\sum_{i\in\mathcal{M}}d_{i}$.
Substituting $z=d_{\mathcal{M}}$ and $z=d_{\mathcal{M}\cup\left\{ u\right\} }$
into $g(z)$, we observe that $g(d_{\mathcal{M}\cup\left\{ u\right\} })-g(d_{\mathcal{M}})=g(\sum_{i\in\mathcal{M}\cup\left\{ u\right\} }d_{i})-g(\sum_{i\in\mathcal{M}}d_{i})=\left(a_{1}-a_{2}\mathrm{e}^{\frac{a_{3}}{D}\sum_{i\in\mathcal{M}\cup\left\{ u\right\} }d_{i}}\right)-\left(a_{1}-a_{2}\mathrm{e}^{\frac{a_{3}}{D}\sum_{i\in\mathcal{M}}d_{i}}\right)<0$
is decreasing and negative due to $d_{\mathcal{M}}<d_{\mathcal{M}\cup\left\{ u\right\} }$
and the monotonicity and concavity of $g(z)$. Additionally, it is
clear that when $\mathcal{M}$ expands, $\sum_{i\in\mathcal{M}}d_{i}b_{i}>0$
is positive and increasing. Therefore, \ding{192} in~(\ref{eq:S_u(mu)})
is proved to be monotonically decreasing. Because $g(z)$ is monotonically
decreasing, it is straightforward to see that \ding{193} in~(\ref{eq:S_u(mu)})
is also monotonically decreasing with the expansion of $\mathcal{M}$.
Finally, we can conclude that $S_{u}(\mathcal{M})$ is monotonically
decreasing, thus proving the submodularity of $S(\mathcal{M})$. 
\end{IEEEproof}
\textcolor{black}{It is worth noting that there is a constraint in
}(\ref{eq:constraint-supply})\textcolor{black}{, also called a knapsack
constraint. This constraint not only affects the resulting social
welfare and the number of the selected miners in the auction, but
also needs a careful auction mechanism design to guarantee the truthfulness.}
Essentially, the optimization problem appears to be a \emph{non-monotone
submodular maximization with knapsack constraints}. It is known that
there is a $\left(0.2-\eta\right)$-approximate algorithm which applies
the fractional relaxation and local search method~\cite[Figure 5]{Lee2009}.
\textcolor{black}{$\eta>0$ is a preset constant parameter that specifies
the approximation ratio $(0.2\lyxmathsym{-}\eta)$.} For the ease
of expression, we name this approximate algorithm as FRLS algorithm.
In general, the FRLS algorithm first solves a linear relaxation of
the original integer problem using local search, and then it rounds
the obtained fractional solution to an integer value. However, the
algorithm requires the objective function to be non-negative. To address
this issue, let $H(\mathcal{M})=S(\mathcal{M})+c\sum_{i\in\mathcal{N}}d_{i}$.
Clearly, $H(\mathcal{M})\geq0$ for any $\mathcal{M}\subseteq\mathcal{N}$
and it remains submodular since $c\sum_{i\in\mathcal{N}}d_{i}$ is
a constant. Additionally, maximizing $S(\mathcal{M})$ is equivalent
to maximizing $H(\mathcal{M})$. Hence, we attempt to design the FRLS
auction which selects the winner based on the FRLS algorithm and let
service price $p_{i}=b_{i}$. As to the specific input to the FRLS
algorithm, it takes $1$ as the number of knapsack constraints, the
normalized demand profile $\frac{\mathbf{d}}{D}$ as its knapsack
weights parameter, $\eta$ as the approximate degree, and $H(\mathcal{M})$
as the value oracle which allows querying for function values of any
given set. The FRLS auction is computationally efficient, as the running
time of the FRLS algorithm is polynomial~\cite{Lee2009}. Furthermore,
miners just need to pay their submitted bids to the CFP and cannot
suffer deficit, so the FRLS auction also satisfies the individual
rationality requirement. However, we find that FRLS auction cannot
guarantee truthfulness. The corresponding proof is omitted due to
space constraints.

\subsection{Multi-Demand miners in Blockchain networks (MDB) Auction }

Although the FRLS auction is capable solving the social welfare maximization
problem approximately, it is not realistic to be directly applied
in a real market since it cannot prevent the manipulation of bids
by bidders, i.e., lacking of truthfulness. As mentioned before, we
aim to design an auction mechanism that not only achieves a good social
welfare, but also possesses the desired properties, including computational
efficiency, individual rationality and truthfulness. Therefore, we
present a novel auction mechanism for Multi-Demand miners in Blockchain
networks (MDB auction). In this auction, the bidders are limited to
be single-minded in the combinatorial auctions. That is, we can assume
safely that the mechanism always allocates to the winner $i$ exactly
the $d_{i}$ items that it requested and never allocates anything
to a losing bidder. The design rationale of the MDB auction relies
on Theorem~\ref{thm:multi-unit-auctrion-truthfulness}. 
\begin{thm}
(\cite{Nisan2015}) In the multi-unit and single minded setting, an
auction mechanism is truthful if it satisfies the following two properties:\label{thm:multi-unit-auctrion-truthfulness} 
\end{thm}
\begin{enumerate}
\item \emph{Monotonicity: If a bidder $i$ wins with bid $(d_{i},b_{i})$,
then it will also win with any bid which offers at least as much price
for at most as many items. That is, bidder $i$ will still win if
the other bidders do not change their bids and bidder $i$ changes
its bid to some $(d_{i}',b_{i}')$ with $d_{i}'\leq d_{i}$ and $b_{i}'\geq b_{i}$. }
\item \emph{Critical payment: The payment of a winning bid $(d_{i},b_{i})$
by bidder $i$ is the smallest value needed in order to win $d_{i}$
items, i.e., the infimum of $b_{i}'$ such that $(d_{i},b_{i}')$
is still a winning bid, when the other bidders do not change their
bids. }
\end{enumerate}

\subsubsection{Auction design}

\begin{algorithm}[tbh]
{\scriptsize{}\scriptsize \begin{algorithmic}[1] \Require{Miners' demand profile~$\mathbf{d}$ and bid profile~$\mathbf{b}$}. \Ensure{Resource allocation $\mathbf{x}$ and service price profile~$\mathbf{p}$.} \Begin 	\ForEach{$i \in \mathcal{N}$} 		\State{$x_i \gets 0, p_i \gets 0$} 	\EndFor
	\State{$\mathcal{M}\gets \varnothing$, $d \gets 0$}
	\While{$\mathcal{M} \neq \mathcal{N}$ } 		\State{$j \gets \arg\max_{i \in \mathcal{N} \setminus \mathcal{M}}S'_{i}(\mathcal{M})$} 		\If{$d + d_j > D$ \textbf{or} $S'_{j}(\mathcal{M}) < 0$} 			\Break 		\EndIf 		\State{$\mathcal{M} \gets \mathcal{M} \cup \{j\}$} 		\State{$d \gets d + d_j$} 	\EndWhile
	\ForEach{$i \in \mathcal{M}$} 		\State{$x_i \gets 1$, $\mathcal{N}_{-i} \gets \mathcal{N} \setminus \{i\}$} 		\State{$\mathcal{T}_0 \gets \varnothing$, $d' \gets 0$, $k \gets 0$,~$L_{p} \gets 0$ } 		\While{$\mathcal{T}_{k} \neq \mathcal{N}_{-i}$} 			\State{$i_{k+1} \gets \arg\max_{l \in \mathcal{N}_{-i} \setminus \mathcal{T}_{k}}S'_{l}(\mathcal{T}_{k})$} 			\State{$b'_{i_{k+1}} \gets \arg_{b_i \in \mathbb{R^+}} S'_{i}(\mathcal{T}_{k})=S'_{i_{k+1}}(\mathcal{T}_{k})$} 			\If{$d' + d_{i_{k+1}} > D$ \textbf{or} $S'_{i_{k+1}}(\mathcal{T}_{k})<0$} 				\Break 			\ElsIf{$d' + d_{i_{k+1}}\leq D - d_i$} 				\State{$L_{p} \gets L_{p} + 1$} 			\EndIf 			\State{$\mathcal{T}_{k+1} \gets \mathcal{T}_{k} \cup \{i_{k+1}\}$, $d' \gets d' + d_{i_{k+1}}$} 			\State{$k \gets k + 1$} 		\EndWhile 		\If{$S'_{i_{L_{p}+1}}(\mathcal{T}_{L_{p}})<0$ \textbf{or} $d_{i_{L_{p}+1}} > d_i$} 			\State{$\widetilde{S} \gets 0$} 		\Else 			\State{$\widetilde{S} \gets S'_{i_{L_{p}+1}}(\mathcal{T}_{L_{p}})$} 		\EndIf 		\State{$b'_{i_{L_p+1}} \gets \arg_{b_i \in \mathbb{R^+}} S'_{i}(\mathcal{T}_{L_p})=\widetilde{S}$}	 		\State{$b'_i \gets \min_{k\in\{0,1,\ldots,L_p+1\}} b'_{i_{k}}$}	 		\State{$p_i \gets (a_{1}-a_{2}\mathrm{e}^{\frac{a_{3}\sum_{j\in\mathcal{M}}d_{j}}{D}})\frac{b'_i}{D}$} 	\EndFor \End \end{algorithmic}\caption{MDB auction\label{alg:BMining-auction}}
}
\end{algorithm}
Before presenting the MDB auction, we first introduce the \emph{marginal
social welfare density}. It is the density of miner $i$'s marginal
social welfare contribution to the existing set of winners $\mathcal{M}$,
which is defined as follows: 
\begin{align}
S_{i}'(\mathcal{M}) & =\frac{S_{i}(\mathcal{M})}{d_{i}}=\frac{S(\mathcal{M}\cup\{i\})-S(\mathcal{M})}{d_{i}}\nonumber \\
 & =\underset{\text{\ding{192}}}{\underbrace{\frac{\left(a_{2}\mathrm{e}^{\frac{a_{3}\sum_{j\in\mathcal{M}}d_{j}}{D}}-a_{2}\mathrm{e}^{\frac{a_{3}\sum_{j\in\mathcal{M}\cup\left\{ i\right\} }d_{j}}{D}}\right)\sum_{j\in\mathcal{M}}d_{j}b_{j}}{Dd_{i}}}}\nonumber \\
 & \qquad+\underset{\text{\ding{193}}}{\underbrace{\left(a_{1}-a_{2}\mathrm{e}^{\frac{a_{3}\sum_{j\in\mathcal{M}\cup\left\{ i\right\} }d_{j}}{D}}\right)\frac{b_{i}}{D}-c}}.\label{eq:marginal_density_SW}
\end{align}
For the sake of brevity, we simply call it \emph{density}.

As illustrated in Algorithm~\ref{alg:BMining-auction}, the MDB auction
allocates computing resources to miners in a greedy way. According
to the density, all miners are sorted in a non-increasing order: 
\begin{equation}
S_{1}'(\mathcal{M}_{0})\geq S_{2}'(\mathcal{M}_{1})\geq\cdots\geq S_{i}'(\mathcal{M}_{i-1})\geq\cdots\geq S_{N}'(\mathcal{M}_{N-1}).\label{eq:density_list}
\end{equation}
The $i$th miner has the maximum density $S_{i}'(\mathcal{M}_{i-1})$
over $\mathcal{N}\setminus\mathcal{M}_{i-1}$ where $\mathcal{M}_{i-1}=\{1,2,\ldots,i-1\}$
and $\mathcal{M}_{0}=\emptyset$. From the sorting, the MDB auction
finds the set of winners $\mathcal{M}_{L_{m}}$ containing $L_{m}$
winners, such that $d_{\ensuremath{\mathcal{M}_{L_{m}}}}\leq D$,
$S_{L_{m}}'(\mathcal{M}_{L_{m}-1})\geq0$ and $S_{L_{m}+1}'(\mathcal{M}_{L_{m}})<0$
(lines~6-13).

To determine the service price for each winner $i\in\mathcal{M}_{L_{m}}$
(lines 14-36), the MDB auction re-executes the winner selection process
and similarly sorts other winners in $\mathcal{N}_{-i}=\mathcal{N}\setminus\{i\}$
as follows: 
\begin{equation}
S_{i_{1}}'(\mathcal{T}_{0})\geq S_{i_{2}}'(\mathcal{T}_{1})\geq\cdots\geq S_{i_{k}}'(\mathcal{T}_{k-1})\geq\cdots\geq S_{i_{N-1}}'(\mathcal{T}_{N-2}),\label{eq:payment_list}
\end{equation}
where $\mathcal{T}_{k-1}$ denotes the first $k-1$ winners in the
sorting and $\mathcal{T}_{0}=\emptyset$. From the sorting, we select
the first $L_{p}$ winners where the $L_{p}$th winner is the last
one that satisfies $S_{i_{L_{p}}}'(\mathcal{T}_{L_{p}-1})\geq0$ and
$d_{\mathcal{T}_{L_{p}-1}}\leq D-d_{i}$. Let $\tilde{S}$ denote
the ($L_{p}+1$)th winner's virtual density. If the ($L_{p}+1$)th
winner has a negative density on $\mathcal{T}_{L_{p}}$, i.e., $S_{i_{L_{p}+1}}'(\mathcal{T}_{L_{p}})<0$,
or its demand is larger than that of winner $i$, i.e., $d_{L_{p}+1}>d_{i}$,
we set $\widetilde{S}=0$. Otherwise, $\widetilde{S}=S_{i_{L_{p}+1}}'(\mathcal{T}_{L_{p}})$.
Meanwhile, Algorithm~\ref{alg:BMining-auction} forms a price list
$\mathbf{L}=\{S_{i_{1}}'(\mathcal{T}_{0}),\ldots,S_{i_{L_{p}}}'(\mathcal{T}_{L_{p}-1}),\widetilde{S}\}$
containing ($L_{p}+1$) density values. According to the list, we
find the winner $i$'s minimum bid $b_{i}'$ such that $S_{i}'(\mathcal{T}_{k-1})\geq S_{i_{k}}'(\mathcal{T}_{k-1}),\exists k\in\{0,1,\ldots,L_{p}\}$
or $S_{i}'(\mathcal{T}_{L_{p}})\geq\widetilde{S}$. Here, $b_{i}'$
is called miner $i$'s ex-ante price, which is the payment without
considering the allocative externalities. Then, we set $p_{i}=\left(a_{1}-a_{2}\mathrm{e}^{\frac{a_{3}\sum_{j\in\mathcal{M}_{L_{m}}}d_{j}}{D}}\right)\frac{b_{i}'}{D}$
as the winner $i$'s final payment.

\subsubsection{Properties of MDB Auction}

We show the computational efficiency (Proposition~\ref{lem:computational_efficient}),
the individual rationality (Proposition~\ref{lem:MDMB_individual_rational}),
and the truthfulness (Proposition~\ref{lem:MDMB_truthful}) of the
MDB auction in the following. 
\begin{prop}
MDB auction is computationally efficient.\label{lem:computational_efficient} 
\end{prop}
\begin{IEEEproof}
In Algorithm~\ref{alg:BMining-auction}, finding the winner with
the maximum density has the time complexity of $O(N)$ (line~7).
Since the number of winners is at most $N$, the winner selection
process (the while-loop lines~6-13) has the time complexity of $O(N^{2})$.
In the service price determination process (lines~14-36), each for-loop
executes similar steps as the while-loop in lines~6-13. Hence, lines~14-36
have the time complexity of $O(N^{3})$ in general. Hence, the running
time of Algorithm~\ref{alg:BMining-auction} is dominated by the
for-loop, which is bounded by polynomial time $O(N^{3})$. 
\end{IEEEproof}
\begin{prop}
MDB auction is individually rational.\label{lem:MDMB_individual_rational} 
\end{prop}
\begin{IEEEproof}
Let $i_{i}$ be the miner $i$'s replacement which appears in the
$i$th place in the sorting~(\ref{eq:payment_list}) over $\mathcal{N}_{-i}$.
Since miner $i_{i}$ would not be in the $i$th place if winner $i$
is considered, we have $S_{i_{i}}'(\mathcal{T}_{i-1})\leq S_{i}'(\mathcal{T}_{i-1})$.
Note that Algorithm~\ref{alg:BMining-auction} chooses the minimum
bid $b_{i}'$ for miner $i$, which means that given the bid $b_{i}'$,
miner $i$'s new density $S_{i}''(\mathcal{T}_{i-1})$ at least satisfies
$S_{i}''(\mathcal{T}_{i-1})\leq S_{i_{i}}'(\mathcal{T}_{i-1})\leq S_{i}'(\mathcal{T}_{i-1})$.
According to the definition of the density in~(\ref{eq:marginal_density_SW}),
$S_{i}'(\mathcal{T}_{i-1})$ is a monotonically increasing function
of $b_{i}$. Hence, we have $b_{i}-b_{i}'\geq0$ as $S_{i}'(\mathcal{T}_{i-1})\geq S_{i}''(\mathcal{T}_{i-1})$.
Therefore, the final payment for miner $i$ is not more than its ex-post
valuation, i.e., $p_{i}=\left(a_{1}-a_{2}\mathrm{e}^{\frac{a_{3}\sum_{j\in\mathcal{M}_{L_{m}}}d_{j}}{D}}\right)\frac{b_{i}'}{D}\leq v_{i}''=\left(a_{1}-a_{2}\mathrm{e}^{\frac{a_{3}\sum_{j\in\mathcal{M}_{L_{m}}}d_{j}}{D}}\right)\frac{b_{i}}{D}$.
Thus, the individual rationality of MDB auction is ensured. 
\end{IEEEproof}
\begin{prop}
MDB auction is truthful.\label{lem:MDMB_truthful} 
\end{prop}
\begin{IEEEproof}
Based on Theorem~\ref{thm:multi-unit-auctrion-truthfulness}, it
suffices to prove that the selection rule of the MDB auction is monotone,
and the ex-ante payment $b_{i}'$ is the critical value for winner
$i$ to win the auction.

We first discuss the monotonicity of the MDB auction in terms of winner
$i$'s bid and demand subsequently. Recalling the density $S_{i}'(\mathcal{M})$
in equation~(\ref{eq:marginal_density_SW}), it is clear that $S_{i}'(\mathcal{M})$
is a monotonically increasing function of miner $i$'s bid $b_{i}$.
As miner $i$ takes the $i$th place in the sorting (\ref{eq:density_list}),
when winner $i$ raises its bid from $b_{i}$ to $b_{i}^{+}$, it
at least has a new larger density $S_{i^{+}}'(\mathcal{T}_{i-1})>S_{i}'(\mathcal{T}_{i-1})\geq0$.
Because of the submodularity of $S(\mathcal{M})$, miner $i$ can
only have a larger density when it is ranked higher in the sorting,
i.e., $S_{i^{+}}'(\mathcal{M}_{i-k})>S_{i^{+}}'(\mathcal{M}_{i-1})\geq0,\forall k\in\{2,3,\ldots,i\}$.
Therefore, miner $i$ with a higher bid can always win the auction.
Similarly, when it comes to miner $i$'s demand $d_{i}$, we only
need to show that $S_{i}'(\mathcal{M})$ is a monotonically decreasing
function of $d_{i}$. Let 
\begin{equation}
h(z)=\frac{a_{4}\left(1-\mathrm{e}^{\frac{a_{3}}{D}z}\right)}{z}\label{eq:h(z)}
\end{equation}
where $z\in\mathbb{R}^{+}$ and all parameters are positive. The first
derivative of $h(z)$ is 
\begin{equation}
\frac{\mathrm{d}h(z)}{\mathrm{d}z}=-\frac{a_{4}(\frac{a_{3}}{D}\mathrm{e}^{\frac{a_{3}}{D}z}z+1-\mathrm{e}^{\frac{a_{3}}{D}z})}{z^{2}}.\label{eq:1st_order_h(z)}
\end{equation}
Since the first derivative of $(\frac{a_{3}}{D}\mathrm{e}^{\frac{a_{3}}{D}z}z+1-\mathrm{e}^{\frac{a_{3}}{D}z})$
is $\frac{a_{3}^{2}}{D^{2}}\mathrm{e}^{\frac{a_{3}}{D}z}z>0$, we
can have $\frac{\mathrm{d}h(z)}{\mathrm{d}z}<0$ with $a_{3},a_{4},D,z>0$.
Thus, $h(z)$ is monotonically decreasing with $z$. By substituting
$z=d_{i}$, we can easily observe that $\text{\ding{192}}$ in~(\ref{eq:marginal_density_SW})
is a monotonically decreasing function with respect to $d_{i}$. Finally,
$S_{i}'(\mathcal{M})$ is proved to be monotonically decreasing with
$d_{i}$ since $\text{\ding{193}}$ in~(\ref{eq:marginal_density_SW})
is clearly a monotonically decreasing function of $d_{i}$ as well.

Next, we prove that $b_{i}'$ is the critical ex-ante payment. This
means that bidding lower $b_{i}^{-}<b_{i}'$ can lead to miner $i$'s
failure in the auction. Given that $d_{i}$ is fixed, we note that
$b_{i}'$ is the minimum bid such that miner $i$'s new density $S_{i}''(\mathcal{T}_{k})$
is no more than any value in the $k$th place in the sorting~(\ref{eq:payment_list}),
where $k\in\{0,1,\ldots,L_{p}-1\}$. If miner $i$ submits a lower
bid $b_{i}^{-}$, it must be ranked after the $L_{p}$th winner in~(\ref{eq:payment_list})
due to submodularity of $S(\mathcal{M})$. Then, its density has to
be compared with $\tilde{S}$. Considering the $(L_{p}+1)$th winner
in the sorting~(\ref{eq:payment_list}), if its density $S_{i_{L_{p}+1}}'(\mathcal{T}_{L_{p}})\geq0$
and $d_{i_{L_{p}+1}}\leq d_{i}$, $\tilde{S}$ is set to be $S_{i_{L_{p}+1}}'(\mathcal{T}_{L_{p}})$.
In this case, miner $i$ with bid $b_{i}^{-}$ cannot take the $(L_{p}+1)$th
place as its new density is $S_{i}''(\mathcal{T}_{L_{p}})<S_{i}'(\mathcal{T}_{L_{p}})\leq\tilde{S}=S_{i_{L_{p}+1}}'(\mathcal{T}_{L_{p}})$.
Also, it no longer can win the auction by taking the place after the
$(L_{p}+1)$th because the remaining supply $D-d_{\mathcal{T}_{L_{p}+1}}$
cannot meet its demand $d_{i}$, i.e., $D-d_{\mathcal{T}_{L_{p}+1}}<d_{i}$.
If $S_{i_{L_{p}+1}}'(\mathcal{T}_{L_{p}})<0$ or $d_{i_{L_{p}+1}}>d_{i}$,
$\tilde{S}$ is just set to be $0$. Apparently, $b_{i}^{-}$ is not
a winning bid as $S_{i}''(\mathcal{T}_{L_{p}})<b_{i}'=\tilde{S}=0$. 
\end{IEEEproof}

\section{Experimental Results and Performance Evaluation\label{sec:Experiment-and-numerical}}

In this section, we first perform experiments to verify the proposed
hash power function and network effects function. Then, from simulation
results, we examine the performance of the proposed auction mechanisms
in social welfare maximization and provide useful decision making
strategies for the CFP and the blockchain developer.

\subsection{Verification for Hash Power Function and Network Effects Function\label{subsec:Verification-for-Hash-Net}}

Similar to the experiments on mobile blockchain mining in~\cite{Xiong2018a,Suankaewmanee2018},
we design a mobile blockchain client application in the Android platform
and implement it on each of three mobile devices (miners). The client
application can not only record the data generated by internal sensors
or the transactions of the mobile P2P data trading, but also allows
each mobile device to be connected to a computing server through a
network hub. The miners request the computing service from the server.
Then, the server allocates the computing resources and starts mining
the block for the miners. At the server side, the each miner's CPU
utilization rate is managed and measured by the Docker platform\footnote{https://www.docker.com/community-edition.}.
In our experiment, all mining tasks (solving the PoW puzzle) are under
Go-Ethereum\footnote{https://ethereum.github.io/go-ethereum.} blockchain
framework. 

\begin{figure}[tbh]
\begin{centering}
\subfloat{\centering{}\includegraphics[width=0.6\columnwidth]{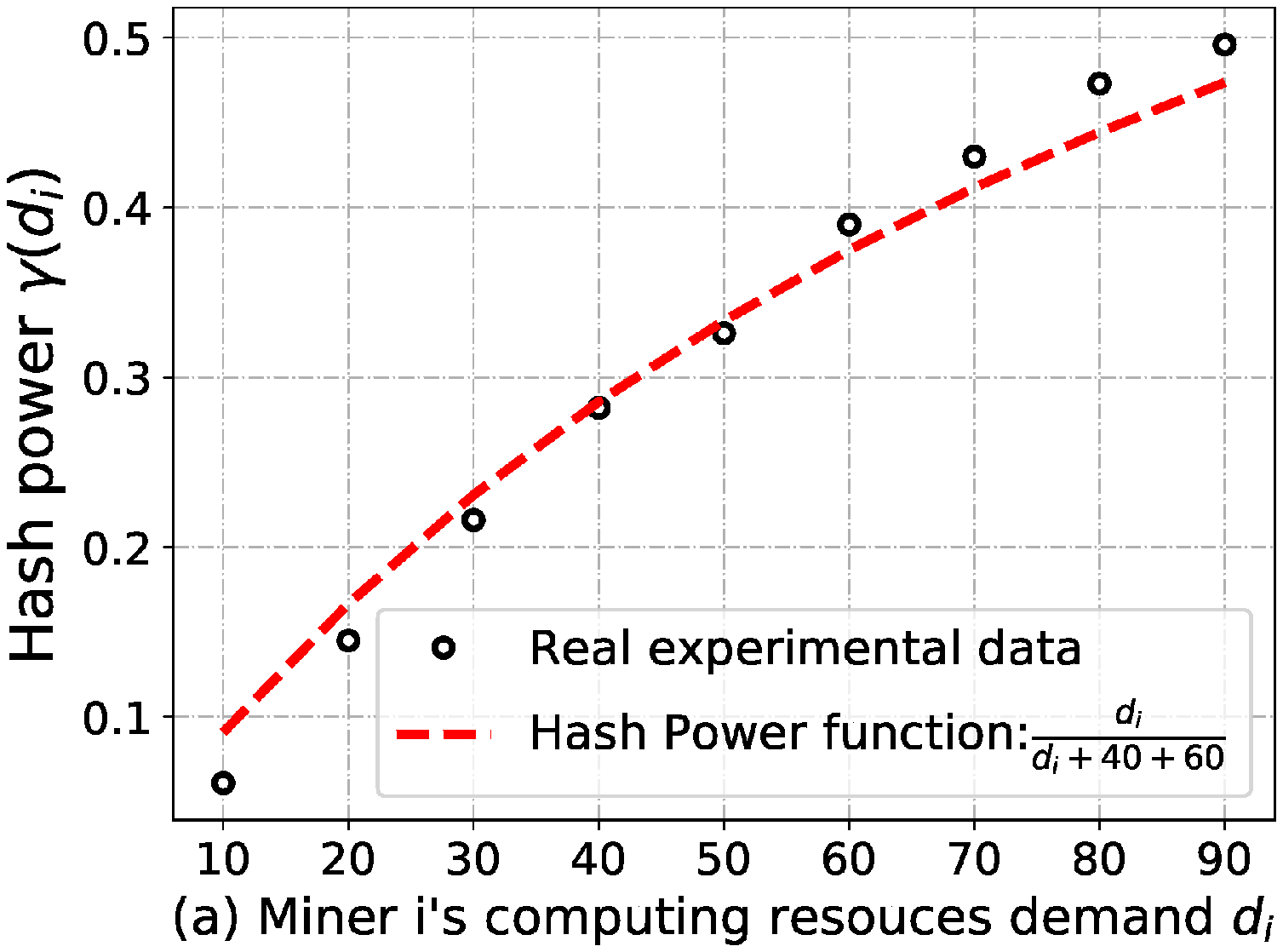} }
\par\end{centering}
\begin{centering}
\subfloat{\centering{}\includegraphics[width=0.6\columnwidth]{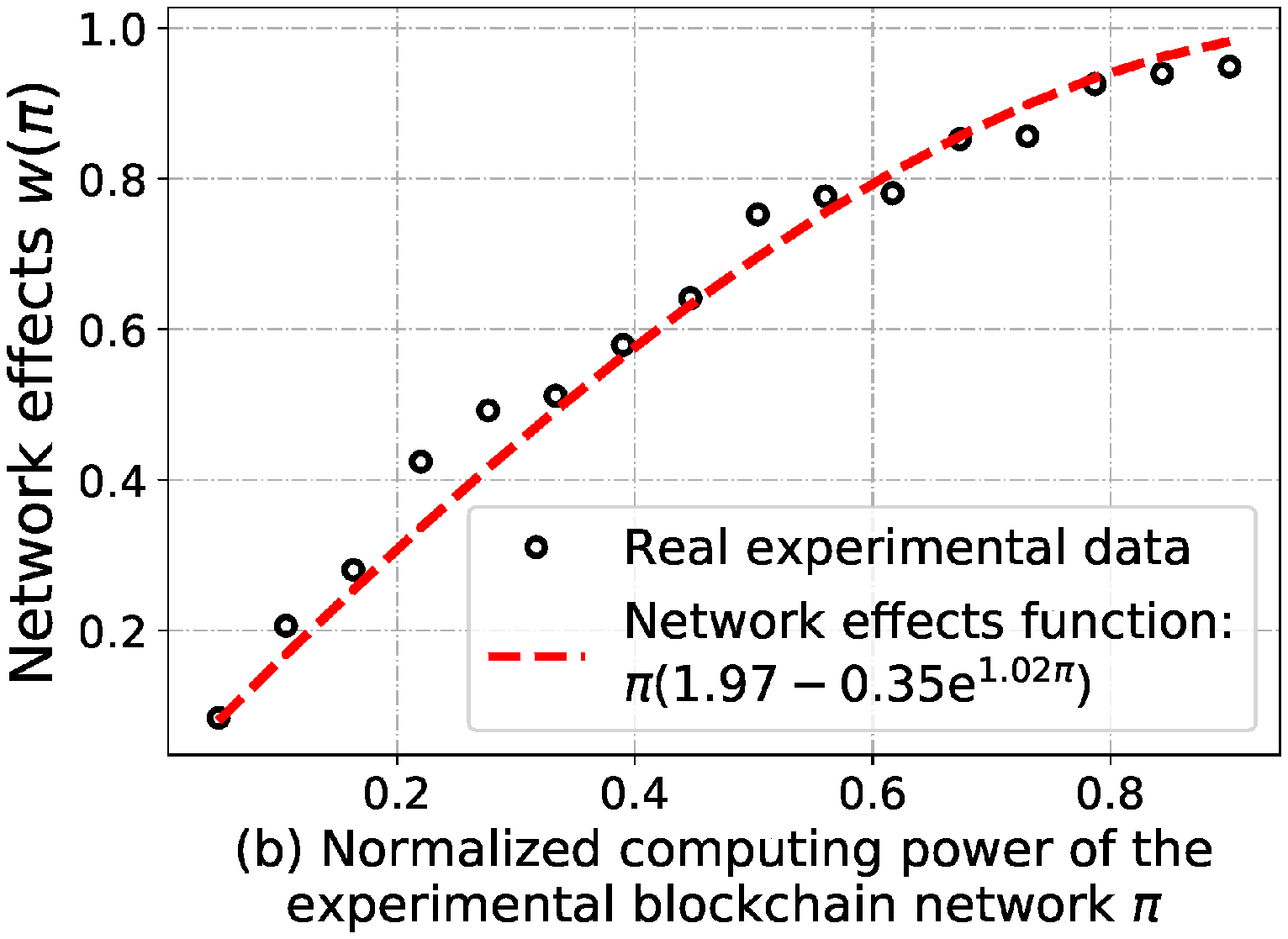} }
\par\end{centering}
\caption{Estimation of (a) the hash power function $\gamma(d_{i})$ in~(\ref{eq:hash-power})
and (b) the network effects function $w(\pi)$ in~(\ref{eq:network_effects_function}).\label{fig:Verification}}
\end{figure}

To verify the hash power function in (\ref{eq:hash-power}), we vary
the service demand of one miner $i$ in terms of CPU utilization,
i.e., $d_{i}$, while fixing the other two miners' service demand
at $40$ and $60$. Here, the total amount of computing resources
is $d_{\ensuremath{\mathcal{N}}}=d_{i}+40+60$. Besides, we initially
broadcast $10$ same transaction records to the miners in the network
so that all mined blocks have the same size. Figure~\ref{fig:Verification}a
shows the change of the hash power, i.e., the probability of successfully
mining a block with different amount of computing resources. We note
that the hash power function defined in (\ref{eq:hash-power}) can
well fit the real experimental results.

To verify the network effects function in~(\ref{eq:network_effects_function}),
we investigate the capability of the blockchain to prevent the double-spending
attacks. We add a malicious miner with fixed computing powers, i.e.,
an attacker performing double-spending attacks, to the blockchain
network. Then, we conduct several tests by varying the CPU resources
of the other miners, i.e., the sum of existing honest miners' computing
resources $d_{\mathcal{N}}$, to measure the probability of the successful
attacks. Specifically we count the number of fake blocks which successfully
join the chain every $10,000$ blocks generated in each test. Based
on the above results, we finally calculate the proportion of the genuine
blocks every $10,000$ blocks (i.e., each data point in the Fig.~\ref{fig:Verification}b)
as the security measure or the network effects of the blockchain network.
As illustrated in Fig.~\ref{fig:Verification}b, it is evident that
the network effects function in~(\ref{eq:network_effects_function})
also well fits the real experiment results. Based on the experiments,
we set $a_{1}=1.97,a_{2}=0.35,a_{3}=1.02$ in the following simulations.

\subsection{Numerical Results}

To demonstrate the performance of the proposed auction mechanisms
and the impacts of various parameters on the social welfare of the
blockchain network, we consider a set of $N$ miners, e.g., mobile
users in a PoW-based blockchain application supported by the CFP.
Each miner's block size is uniformly distributed over $(0,1024]$.
Instead of being restricted to submit a constant demand as in the
CDB auction, each miner in the MDB auction and FRLS auction can choose
its desired demand which follows the uniform distribution over $[{\beta_{1}}D,{\beta_{2}}D]$.
Except Fig. 6a, each measurement is averaged over $600$ instances
and the associated $95$\% confidence interval is given. We can find
that the confidence intervals are very narrowly centered around the
mean. The default parameter values are presented in Table~\ref{tab:Default-Parameter-Values}.
Note that setting $q=10$, $\beta_{1}=0$ and $\beta_{2}=0.02$ means
the expected demand of miners in the MDB auction is equal to the constant
demand of miners in the CDB auction. Hence, we can compare the performance
of both proposed auction mechanisms.
\begin{table}[tbh]
\centering{}

\caption{Default parameter values\label{tab:Default-Parameter-Values}}

\centering{}%
\begin{tabular}{|c|c||c|c|}
\hline 
Parameters  & Values  & Parameters  & Values\tabularnewline
\hline 
\hline 
$N$  & $300$  & $T$  & $12.5$\tabularnewline
\hline 
$r$  & $0.007$  & $\lambda$  & $15$\tabularnewline
\hline 
$c$  & $0.001$  & $q$  & $10$\tabularnewline
\hline 
$a_{1}$  & $1.97$  & $\beta_{1}$, $\beta_{2}$  & $0$, $0.02$\tabularnewline
\hline 
$a_{2}$  & $0.35$  & $\xi$  & $0.001$\tabularnewline
\hline 
$a_{3}$  & $1.02$  & $D$  & $1000$ \tabularnewline
\hline 
\end{tabular}
\end{table}
\begin{table}[tbh]
\begin{centering}
\caption{MDB auction versus FRLS auction in social welfare maximization\label{tab:FRLS-versus-MDMB}}
\begin{tabular}{|c|c|c|c|c|}
\hline 
Number of miners  & $10$  & $15$  & $20$  & $25$\tabularnewline
\hline 
\hline 
MDB auction  & $33.954$  & $50.368$  & $65.421$  & $80.135$\tabularnewline
\hline 
FRLS auction  & $34.656$  & $49.935$  & $65.060$  & $79.853$\tabularnewline
\hline 
\end{tabular}
\par\end{centering}
\end{table}

\subsubsection{Evaluation of MDB auction versus FRLS auction in terms of social
welfare maximization}

We evaluate the performance of the MDB auction in maximizing the social
welfare by comparing it with the FRLS auction. Table~\ref{tab:FRLS-versus-MDMB}
shows the social welfare obtained by the MDB auction and the FRLS
auction. The social welfare generated from the MDB auction is lower
than that from the FRLS auction when dealing with a small number of
miners. As the group of interested miners grows, the MDB auction can
achieve slightly larger social welfare although it has to preserve
necessary economic properties, including individual rationality and
truthfulness. The main reason is that the FRLS auction is an algorithm
which only provides a theoretical lower bound guarantee in the worst
case for approximately maximizing the social welfare, and may have
more severe performance deterioration when interested miners become
more. 

\subsubsection{Impact of the number of miners $N$}

\begin{figure}[tbh]
\begin{centering}
\includegraphics[width=1\columnwidth]{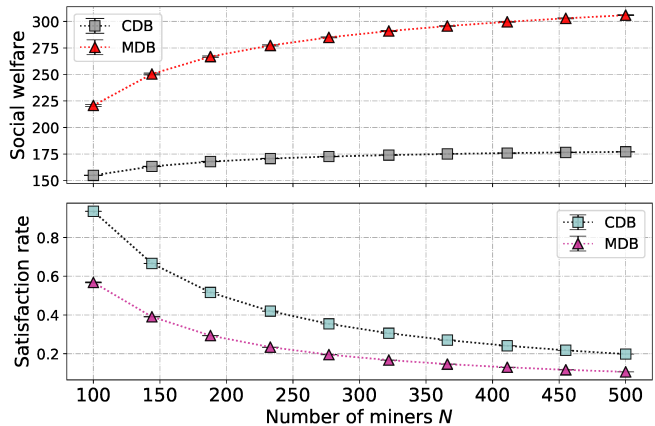}
\par\end{centering}
\centering{}\caption{Impact of the number of miners $N$. \label{fig:N}}
\end{figure}
Besides the social welfare, we introduce the satisfaction rate, i.e.,
the percentage of winners selected from all interested miners, as
another metric. Here, we compare the social welfare as well as the
satisfaction rate of the CDB auction and the MDB auction with various
number of miners, as shown in Fig.~\ref{fig:N}. From Fig.~\ref{fig:N},
we observe that the social welfare $S$ in both auction mechanisms
increases as the base of interested miners becomes larger. We observe
that the satisfaction rate decreases and the rise of the social welfare
also slows down with the increase of $N$. The main reason is that
the competition among miners becomes more obvious when more miners
take part in the auction, and, with more winners selected by the auction,
the subsequent winner's density decreases due to the network effects.
When choosing between the CDB auction and the MDB auction, Fig.~\ref{fig:N}
clearly shows that there is a tradeoff between the social welfare
and the satisfaction rate. The MDB auction can help the CFP achieve
more social welfare than the CDB auction because of its advantage
in relaxing restrictions on miners' demand. However, the CDB is relatively
more fair because the MDB auction allows miners with large demand
to take up more computing resources and this leads to a lower satisfaction
rate.

\subsubsection{Impact of the unit cost $c$, the fixed bonus $T$, the transaction
fee rate $r$ and the block time $\lambda$}

\begin{figure}[tbh]
\begin{centering}
\includegraphics[width=1\columnwidth]{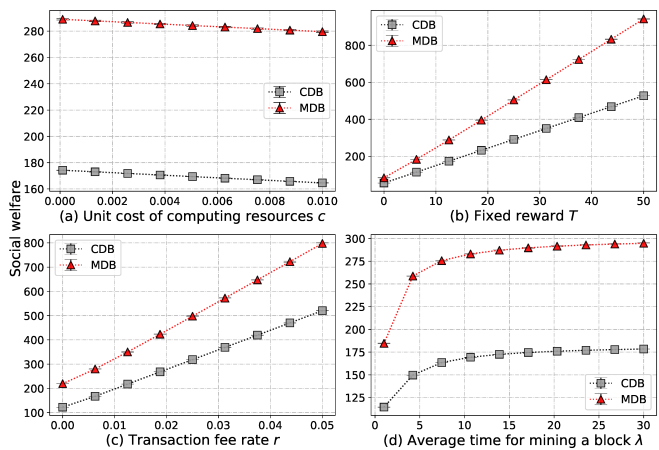}
\par\end{centering}
\centering{}\caption{Impact of unit cost $c$, fixed bonus $T$, transaction fee rate $r$
and block time $\lambda$. \label{fig:cTrl}}
\end{figure}
The CFP organizes the auction and cares about the unit cost of the
computing resource. It is obvious from Fig.~\ref{fig:cTrl}~(a)
that as the computing resources become expensive, the social welfare
in each auction mechanism decreases linearly. The blockchain developer
may be more interested in optimizing the blockchain protocol parameters,
including the fixed reward, the transaction fee rate and the block
time. In Figs.~\ref{fig:cTrl}(b)-(d), we study their impacts on
the social welfare of the blockchain network. Figures.~\ref{fig:cTrl}(b)~and~\ref{fig:cTrl}(c)
illustrate that if the blockchain developer raises the fixed bonus
$T$ or the transaction fee rate $r$, higher social welfare will
be generated nearly in proportion. This is because miner's valuation
increases with higher $T$ and $r$, according to the definition in~(\ref{eq:ex-ante}).
Moreover, by increasing $T$ and $r$, we observe that the difference
of the social welfare between the CDB auction and the MDB auction
amplifies. The reason is that raising $T$ and $r$ can significantly
improve the valuation of miner $i$ which possesses large block size
$s_{i}$ and high demand $d_{i}$. As shown in Fig.~\ref{fig:cTrl}~(d),
when the blockchain developer raises the difficulty of mining a block,
i.e., extending the block time $\lambda$, the social welfare goes
up. This is because a long block time $\lambda$ gives the miner which
has solved the PoW puzzle a higher probability to successfully propagate
the new block and reach consensus. However, different from adjusting
$T$ and $r$, the marginal gains in social welfare gradually become
smaller if the blockchain developer continues to increase the difficulty
of the blockchain mining. This is mainly due to that the increasing
value of $\lambda$ has less impact on the miner's valuation, as can
be seen from the equations~(\ref{eq:p-mining-a-block}) and~(\ref{eq:ex-ante}).
\begin{figure}[tbh]
\begin{centering}
\includegraphics[width=1\columnwidth]{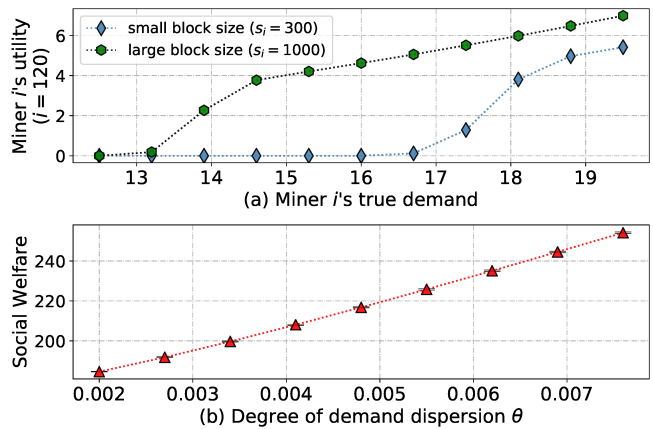}
\par\end{centering}
\centering{}\caption{Relationship between miner $i$'s ($i=120$) utility and its true
demand, and the impact of the degree of demand dispersion $\theta$.\label{fig:DDD}}
\end{figure}

\subsubsection{Miner's utility and individual demand constraints in the MDB auction}

In the MDB auction, we randomly choose a miner (ID=$120$) to see
its utility which is defined by the difference between its ex-post
valuation and its payment, i.e., $v''_{120}-p_{120}$. The miner's
block size is respectively at a low level ($s_{120}=300$) and a high
level ($s_{120}=1000$). We investigate the impact of the miner's
true demand on its utility, which also reflects the impact of its
available budget. Fig.~\ref{fig:DDD} (a) shows that when miner 120\textquoteright s
true demand rises, its utility initially stays at $0$ and then suddenly
increases. This indicates that only when the miner's demand is above
a threshold, it can be selected as the winner by the MDB auction,
i.e., $x_{i}$ changes immediately from $0$ to $1$, obtains the
computing resources and finally has a positive utility. Otherwise,
the miner would not be allocated the resources, i.e., $x_{i}=0$,
and then both its ex-post valuation and payment should be $0$ according
to the MDB auction algorithm, which results in zero utility. Additionally,
if the miner's generated block is larger, it can obtain higher utility
with the same true demand. This implies that miners with large block
size and high demand are easier to be selected by the MDB auction
for social welfare maximization. 

In Fig.~\ref{fig:DDD} (b), we investigate the impact of the demand
constraints on the social welfare in the MDB auction. To fix the miner's
expected demand at $q$, we set demand constraints $\beta_{1}D=q-{\theta}D$
and $\beta_{2}D=q+{\theta}D$ where ${\theta}\in[0,\min(\frac{q}{D},1-\frac{q}{D})]$
characterizes the degree of demand dispersion. It is clear that social
welfare increases as the degree of demand dispersion rises and miners
have more freedom to submit their desired demands.

\section{Conclusions\label{sec:Conclusions}}

In this paper, we have investigated the cloud/fog computing services
that enable blockchain-based DApps. To efficiently allocate computing
resources, we have presented an auction-based market model to study
the social welfare optimization and considered allocative externalities
that particularly exist in blockchain networks, including the competition
among the miners as well as the network effects of the total hash
power. For miners with constant demand, we have proposed an auction
mechanism (CDB auction) that achieves optimal social welfare. For
miners with multiple demands, we have transformed the social welfare
maximization problem to a non-monotone submodular maximization with
knapsack constraints problem. Then, we have designed two efficient
mechanisms (FRLS auction and MDB auction) maximizing social welfare
approximately. We have proven that the proposed CDB and MDB auction
mechanisms are truthful, individually rational and computationally
efficient and are able to solve the social welfare maximization problem. 

\textcolor{black}{In this work, we consider the energy and computational
constraints for PoW-based public blockchain network while assuming
an ideal communication environment. For practical system implementation,
the communication constraint is actually an important factor in establishing
the mobile blockchain network. An example is that the limited bandwidth
for each miner's mutual wireless communication will not only affect
each miner's utility, but also have an adverse impact on the block
broadcasting process and the throughput of the whole blockchain network.
For the future work, we will take the complicated communication environment
into account, and design new spectrum allocation algorithms for more
efficient and practical blockchain system.}

\bibliographystyle{ieeetr}
\bibliography{PJ4_CFP_Blockchain}

\end{document}